# The Nature of Intelligence

Barco Jie You[1,*]

[1]Institute for Computer Science, Heidelberg University, Heidelberg, 69120, Germany
[*]barcojie@gmail.com

## Abstract

The human brain is the substrate for human intelligence. By simulating the human brain, artificial intelligence builds computational models that have learning capabilities and perform intelligent tasks approaching the human level. Deep neural networks consist of multiple computation layers to learn representations of data and improve the state-of-the-art in many recognition domains. However, the essence of intelligence commonly represented by both humans and AI is unknown. Here, we show that the nature of intelligence is a series of mathematically functional processes that minimize system entropy by establishing functional relationships between datasets over the space and time. Humans and AI have achieved intelligence by implementing these entropy-reducing processes in a reinforced manner that consumes energy. With this hypothesis, we establish mathematical models of language, unconsciousness and consciousness, predicting the evidence to be found by neuroscience and achieved by AI engineering. Furthermore, a conclusion is made that the total entropy of the universe is conservative, and the intelligence counters the spontaneous processes to decrease entropy by physically or informationally connecting datasets that originally exist in the universe but are separated across the space and time. This essay should be a starting point for a deeper understanding of the universe and us as human beings and for achieving sophisticated AI models that are tantamount to human intelligence or even superior. Furthermore, this essay argues that more advanced intelligence than humans should exist if only it reduces entropy in a more efficient energy-consuming way.

## Introduction

Artificial intelligence (AI) is the field of computer science that attempts to build machines that can think, learn and act intelligently, similar to a human. AI systems are powered by machine learning algorithms and neural networks, which allow them to learn directly from massive amounts of data. AI has progressed rapidly due to advances in deep learning and neural networks[1], increases in data and computing power, and innovative application of algorithms for reinforcement learning[2]. With the remarkable progress of generative AI[3], machines continue to match and exceed human-level performance on an expanding set of cognitive tasks, spurring excitement and discussion about the future potential impacts of AI on jobs, society and the world. The pace of progress shows no signs of slowing, indicating that new life with reliable and capable AI may arrive sooner than previously anticipated. Therefore, what is the nature of intelligence, and can AI develop into a form of superintelligence?

ChatGPT[4], as a chatbot, shows emergent properties to some extent while communicating with humans. From the perspective of Turing Test[5], has machine passed (intentionally or unconsciously) a human? While it is difficult to answer this question in an objective manner, it is phenomenally definite that AI systems have shown immense cognitive capabilities. Can these phenomena shed light on the essence of human intelligence regarding unconsciousness as well as consciousness?

Throughout Earth's history, life has exhibited a tendency towards increasing order and intelligence over time through the process of evolution by natural selection. The environment on Earth to support life exhibited a similar tendency towards increasing order before and after the emergence of life, finally leading to the birth of human intelligence, which in turn created artificial intelligence. Are there any common patterns formulated from or a common nature regulating the development of intelligence since the very beginning of the universe?

To address the above questions, this essay dives into the mathematical nature of contemporary AI systems, reviews what is known about the developmental process of biological intelligence and the establishment of dependably habitable conditions for biological intelligence, and furthermore proposes a novel hypothesis about the theoretical framework of intelligence, arguing that intelligence is a series of mathematically functional processes that minimize system entropy by establishing functional relationships between datasets over the space and time. Based on this hypothesis, mathematical models of human language, unconsciousness and consciousness are established, with predictions that could be verified by experimental researches in relevant areas, giving insights into how to build conscious machines.

This paper concludes that intelligence is a series of parallel processes to spontaneous processes that comply with the second law of thermodynamics. The intelligent processes counter the spontaneity of entropy increase with energy release, but reduce entropy by consuming energy in a self-reinforced manner. Furthermore, datasets, which are two essential components of an intelligent process as input and output, are all pre-existing in the universe, while intelligence has the mission to establish functional connections between the datasets physically or informationally across the space and time.

## Deep Learning

Deep learning is a representation learning method that allows a machine to be fed with raw data and to automatically discover the representations needed for cognitive tasks[1]. A deep neural network consists of multiple layers of units, also called neurons, each representing a nonlinear function $\theta$ ($\theta \in \boldsymbol{\theta}$) that transforms the representation at one level (starting with the raw input $\boldsymbol{X}$) into a representation at a higher, slightly more abstract level until a final layer represents the output $\boldsymbol{Y}$ (Fig. 1). By stacking enough such transformation layers, a deep neural network can represent complex functions. Thanks to the back-propagation technique[6], the parameters of functions $\boldsymbol{\theta}$ representing the units of a multilayer neural network can be learned through gradient descent[7,8], providing there is a set of samples that include inputs $\{\boldsymbol{x_i}\}$ and corresponding outputs $\{\boldsymbol{y_i}\}$ (Fig. 2). This procedure is called neural network training. As shown in Fig. 2, the neural network maps an input $\boldsymbol{x_i}$ to an output $\hat{\boldsymbol{y}}_i$ via a feedforward pass, and an error between the prediction $\hat{\boldsymbol{y}}_i$ and actual output $\boldsymbol{y_i}$ will be sent back to each neuron in the backward direction, with which the derivative of the error (gradient) with respect to each parameter (or weight) of the modular function ($\theta$) is calculated; thus, an optimization to reduce the errors by adjusting the parameters can be achieved.

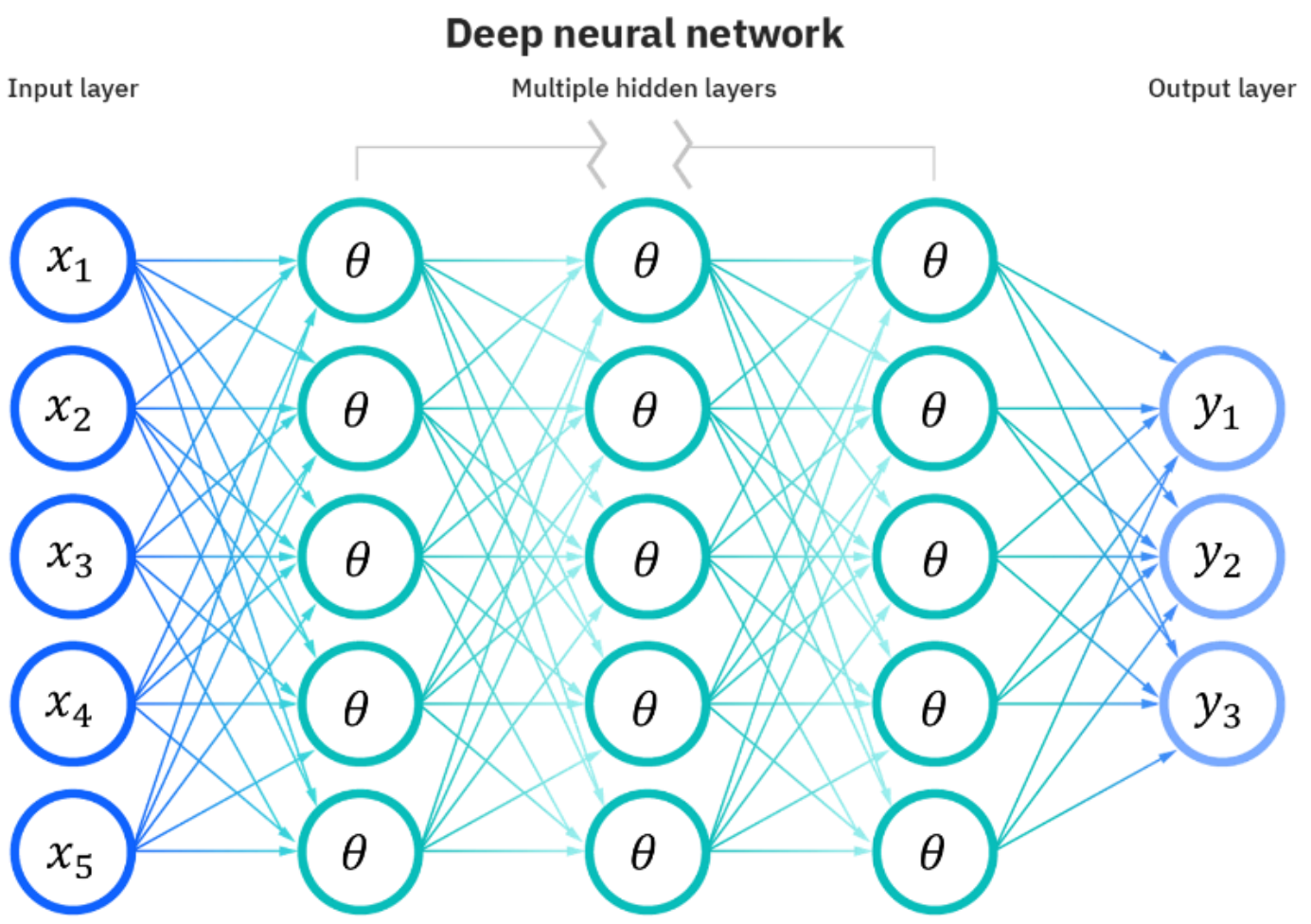

**Figure 1. Neural network for deep learning.** A multilayer neural network shown by connected nodes (circles), with the input layer feeding in scalar elements of a random variable and output layer emitting out scalar elements of another random variable. Variable hidden layers consist of nodes representing modular functions ($\theta \in \boldsymbol{\theta}$) that take inputs from the last layer and output to the next layer.

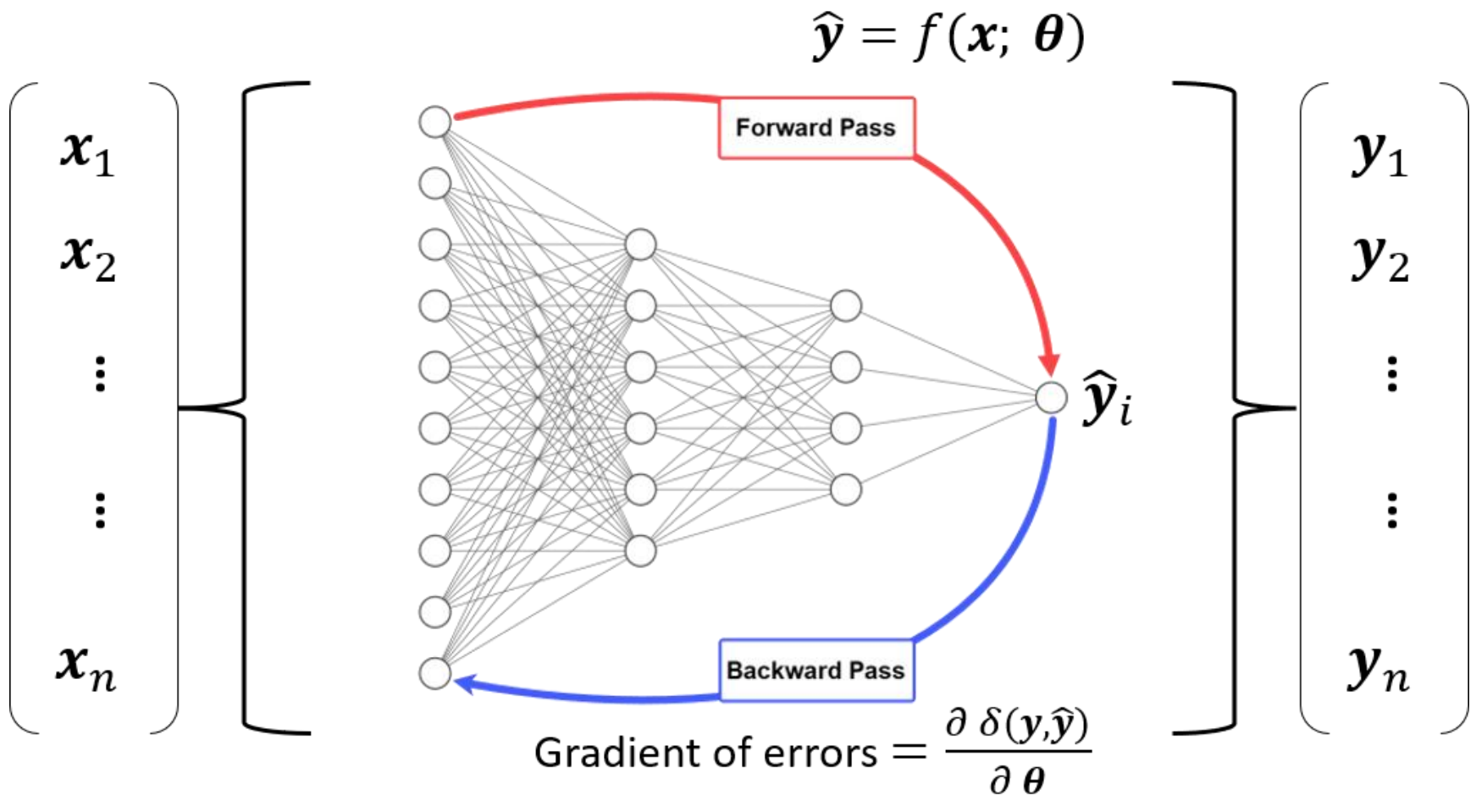

**Figure 2. Feedforward pathway and backpropagation in a multilayer neural network.** A multilayer network is trained by a sample dataset through the feedforward pathway and backpropagation of gradient descent. Sample values of variable $\boldsymbol{X}$ ($\{\boldsymbol{x}_i\} \in \boldsymbol{X}$) are transformed layer by layer in the forward pass to prediction values $\{\hat{\boldsymbol{y}}_i\}$, and then errors between prediction values and actual sample values of $\boldsymbol{Y}$ are calculated, with their gradients with respect to the modular function parameters ($\frac{\partial \delta}{\partial \boldsymbol{\theta}}$) descending towards 0 in a backpropagated way.

Without loss of generality and accuracy, the working principle of a deep neural network can be expressed as equation (1), in which function $f$ represents the architecture of the neural network, $\boldsymbol{\theta}$ represents a set of modular functions (parameters) determining each neuron's computational behaviour, and an input $\boldsymbol{X}$ is mapped to an output $\boldsymbol{Y}$ by function $f$ in case the parameters $\boldsymbol{\theta}$ are certain. If there is a set of samples $(\{\boldsymbol{x}\}, \{\boldsymbol{y}\})$, and in case the structure of function $f$ is determined, the set of parameters $\boldsymbol{\theta}$ can be inferred (learned) through gradient descent and backpropagation methods. According to the state-of-the-art research in AI, the structure of each modular function $\theta$ should be certain in a deep neural network, and the function $f$ should be differentiable with respect to its input and parameters.

$$\boldsymbol{Y} \leftarrow f(\boldsymbol{X}; \boldsymbol{\theta}) \qquad (1)$$

Then, the deep learning problem is simplified as an engineering procedure to seek a function $f(\cdot\,; \boldsymbol{\theta})$ that minimizes the errors between estimates $\widehat{\boldsymbol{Y}} = f(\boldsymbol{X}; \boldsymbol{\theta})$ and actual $\boldsymbol{Y}$ based on an existing dataset mapping: $\{\boldsymbol{x}\} \rightarrow \{\boldsymbol{y}\}$. For example, to build a model that can classify images, we first need to collect a large set of images of various objects and label these images with meaningful tags (symbols, words, or anything else, which can uniquely identify the categories of objects). Every image is an input $\boldsymbol{x}$, and the corresponding label is an output $\boldsymbol{y}$ (Fig. 3). With this set of images and labels, a machine learning model $f(\cdot\,; \boldsymbol{\theta})$ can be found towards minimizing the prediction errors, and this is called supervised learning. In the practice of AI engineering, the image is expressed by a matrix with multiple channels (each channel represents pixels in a colour) and the label is expressed by a vector. Then, the error between an estimate and actual output is expressed by a mathematical distance between vectors, which is also called the cost/loss function. Designing or seeking proper cost functions $\delta(\cdot)$ is also a major task for contemporary AI engineering.

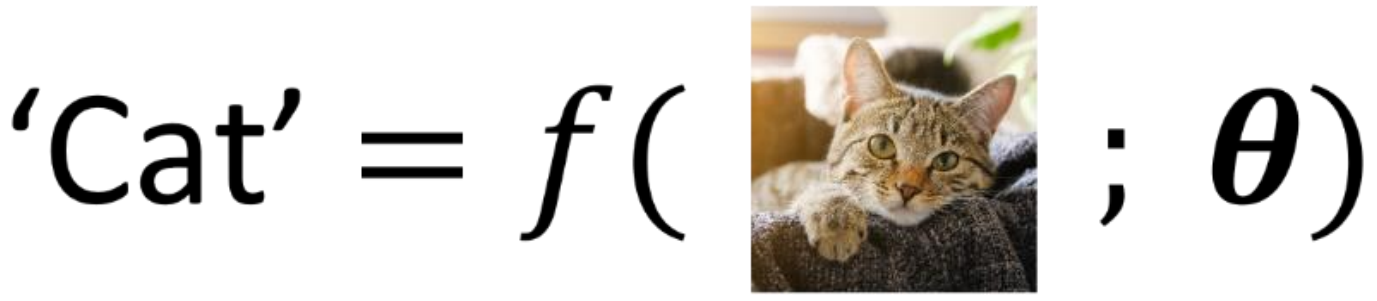

**Figure 3. Schematic diagram of mapping an image to a category.** An image is expressed as a tensor with channels to represent pixels in colour planes. As an input variable, an image can be transformed to a categorical value represented by an English word.

In summary, a supervised learning procedure, deep or not, can be formulated as equation (2).

$$\begin{cases} \boldsymbol{Y} \quad \leftarrow \quad f(\boldsymbol{X}; \boldsymbol{\theta}) \\ \delta(f(\boldsymbol{X}; \boldsymbol{\theta}), \boldsymbol{Y}) \rightarrow 0 \end{cases} \qquad (2)$$

## Reinforcement Learning

To achieve a function in equation (2) as in supervised learning, there should be samples exemplifying the ‘ground-truth’ mapping between inputs and outputs and a ‘ground-truth’ judging standard $\delta(\cdot)$. Nevertheless, in practical applications, both of the above conditions are missing. Imagining in the early ages of human evolution, there was no right from wrong.

Reinforcement learning is a machine learning approach where agents learn how to achieve a goal in a complex, uncertain environment[2]. Reinforcement learning allows machines and software agents to automatically determine the ideal behaviour within a specific context to maximize performance. It is a trial-and-error learning method that does not rely on exemplars to show the optimal solution. The agent interacts with the environment, chooses actions, observes rewards and new states, learns from these experiences, and continually adjusts its strategy to gain the maximum cumulative reward. The agent is not told which actions to take, but it must discover them itself based on the effects of its own decisions. With time and experience, the agent becomes adept at achieving its goal in the environment.

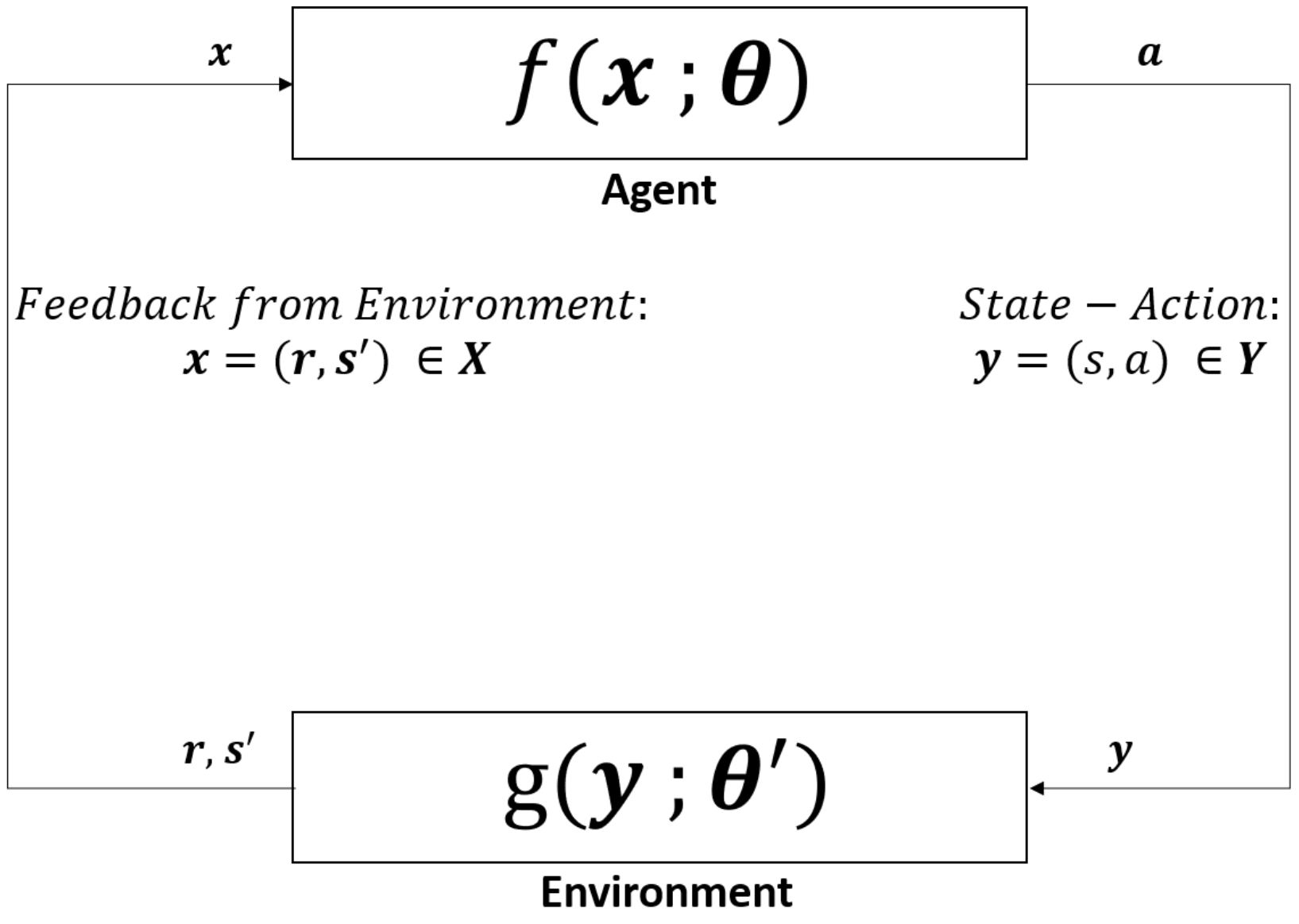

**Figure 4. The agent-environment interaction in a reinforcement learning process.** An agent represented by function $f(\cdot)$ receives inputs ($\boldsymbol{X}$) from the environment, which is represented by another function $g(\cdot)$ conceived by the agent. $\boldsymbol{X}$ is a composed variable by combining the currently perceived environment state ($\boldsymbol{S}$) and reward ($\boldsymbol{R}$) given by the environment at the last time step, which is transformed by $f(\cdot)$ into an action ($\boldsymbol{a}$) performed by the agent upon the environment. Then, $g(\cdot)$ transforms the current state ($\boldsymbol{s}$) and action ($\boldsymbol{a}$) into a new state ($\boldsymbol{s}'$), rewarding $\boldsymbol{r}$. Over time, the agent interacts with the environment in a continuous way or in episodes to minimize the errors of the $g(\cdot)$ function's predictions to targets. In reinforcement learning, $f(\cdot)$ is called the policy, and $g(\cdot)$ is called the value function. Both parameters are optimized by gradient descent over loss functions derived from the divergence between actual returns and estimated values, which is also a value variance reduction process based on the Bellman equation. The gradients flow from value function $g(\cdot)$ to policy $f(\cdot)$.

In reinforcement learning, an agent relies on the environment to provide an evaluation of its behaviour to map an input (a reward $\boldsymbol{r}$ and environment's current state $\boldsymbol{s}$) to an action $\boldsymbol{y}$ (Fig. 4). By mathematically combining reward $\boldsymbol{R}$ and environment's state $\boldsymbol{S}$ into a variable $\boldsymbol{X}$ and representing the state-action pair as $\boldsymbol{Y}$, equation (1) is still applied for a reinforcement-learning agent. The feedback provided by the environment lets the agent know if it achieves the goal or acts desirably. The agent uses the rewards to learn which actions in each state are optimal. Therefore, reinforcement learning techniques seek to define proper goals and evaluate the actions of agents towards achieving goals. Several practical methods for agents to learn from interacting with the environment have been developed, such as temporal difference (TD), actor-critic and Q-learning[2]. Whichever method, an agent needs to estimate a value based on the current situation (received reward $\boldsymbol{R}$, action performed $\boldsymbol{A}$, and environment state $\boldsymbol{S}$). Representing the agent and environment as deep neural networks, the parameters of modular functions can be learned through gradient descent and backpropagation. A reinforcement learning problem can be expressed as equation (3), in which function $f$ represents agent, $\boldsymbol{X}$ represents perceived circumstances, $\boldsymbol{Y}$ represents actions performed by agent and relevant environment states, function $g$ represents the environment model conceived by the agent for evaluating its acts, $\boldsymbol{V}$ represents the value (immediate or long-term) estimated by the agent according to its conceived environment model, and $\delta$ represents the errors between estimated values and actual returns received from true environment. Then, the function parameters are optimized by a value-variance minimization process based on Bellman equation[9] over gradient descent, and gradients are propagated from $g$ to $f$.

$$\begin{cases} \boldsymbol{Y} \quad \leftarrow \quad f(\boldsymbol{X};\ \boldsymbol{\theta}) \\ \boldsymbol{V} \quad \leftarrow \quad g(\boldsymbol{Y};\ \boldsymbol{\theta}') \\ \delta(g(\boldsymbol{Y};\boldsymbol{\theta}'),\boldsymbol{V}) \rightarrow 0 \end{cases} \qquad (3)$$

## Generative AI

Generative AI allows machines to generate new examples from scratch rather than labelling or classifying existing examples[3]. Various methods, such as GANs[10], VAEs[11], autoregressive models[12] and transformers[13], can generate images, audio, text, video and more. Generative techniques are useful for data augmentation, simulation, and tasks requiring human-level creativity by applying unsupervised or self-supervised machine learning to a dataset. Generative AI continues to push the boundaries of what is possible with machine intelligence.

Generative adversarial networks (GANs) formulate the unsupervised learning problem as a game between two opponents: a generator $f(\cdot\,;\boldsymbol{\theta}^{(G)})$ that samples from a distribution and a discriminator $g(\cdot\,;\boldsymbol{\theta}^{(D)})$ that classifies the samples as real or false. Typically, the generator is represented as a deterministic feedforward neural network through which a fixed noise source $\boldsymbol{Z}\sim N(0,I)$ is passed, and the discriminator is another neural network that maps an image to a binary classification probability. The GANs game is then formulated as a zero-sum game where the value is the cross-entropy loss between the discriminator's prediction and the identity of the image as real or generated, which is minimized with respect to the parameters of the discriminator ($\boldsymbol{\theta}^{(D)}$) and maximized with respect to the parameters of the generator ($\boldsymbol{\theta}^{(G)}$), to finally reach a Nash equilibrium[10] (Fig. 5).

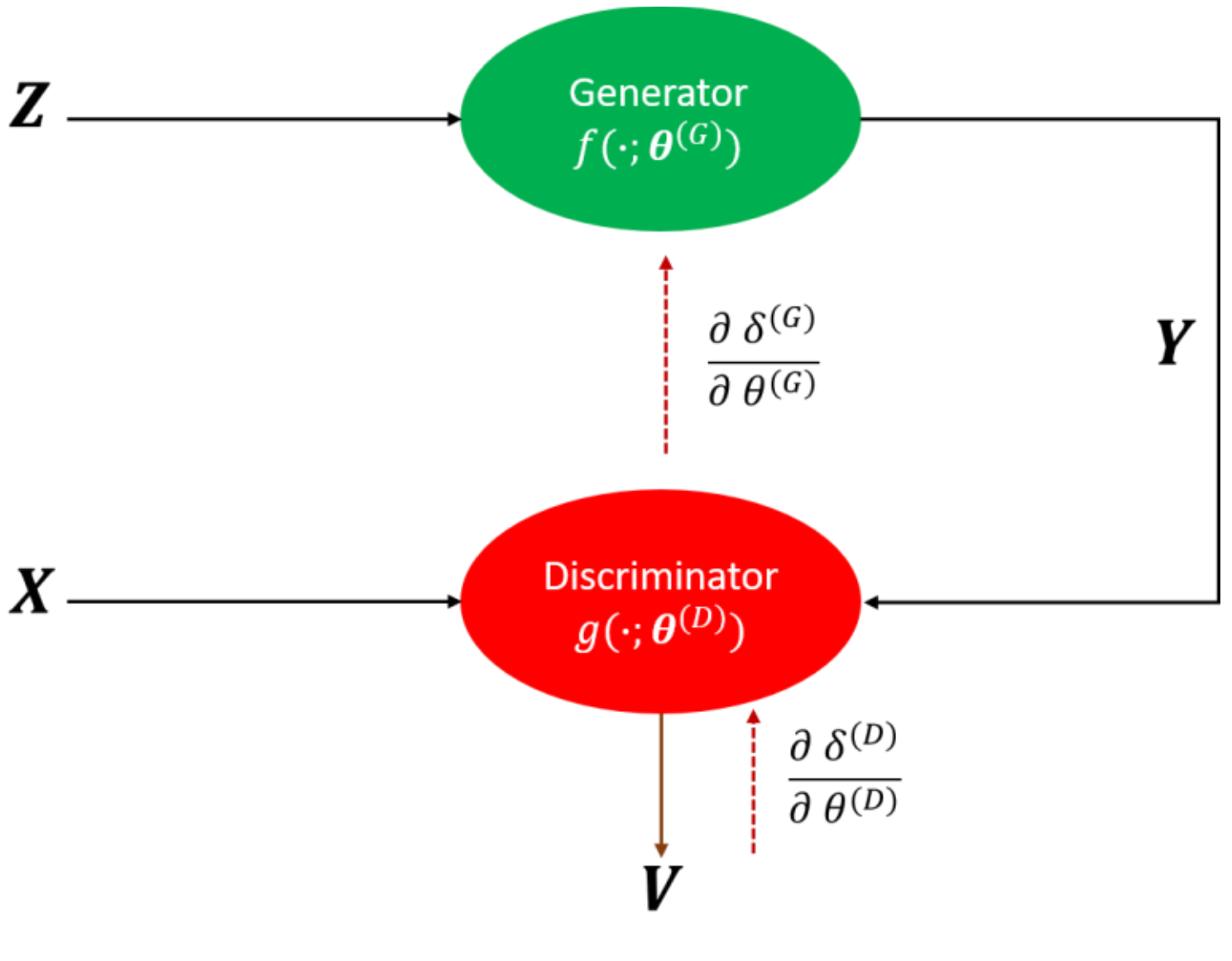

**Figure 5. Information structure of GANs and gradient flow paths.** GANs are composed of two functions, each of which is differentiable with respect to both its inputs and parameters. The generator is a function $f(\cdot;\boldsymbol{\theta}^{(G)})$ that takes a random variable $\boldsymbol{Z}$ as input and $\boldsymbol{\theta}^{(G)}$ as parameters, while the discriminator is a function $g(\cdot;\boldsymbol{\theta}^{(D)})$ that takes samples $\boldsymbol{X}$ as input and $\boldsymbol{\theta}^{(D)}$ as parameters. Both components have loss functions defined in terms of both their parameters, as $\delta^{(G)}(\boldsymbol{\theta}^{(G)},\boldsymbol{\theta}^{(D)})$ and $\delta^{(D)}(\boldsymbol{\theta}^{(G)},\boldsymbol{\theta}^{(D)})$, respectively, and both wish to minimize their losses by controlling their own parameters because they cannot control others' parameters. The optimization of $(\boldsymbol{\theta}^{(G)},\boldsymbol{\theta}^{(D)})$ is to reach a Nash equilibrium, obtaining a local minimum of $\delta^{(G)}$ with respect to $\boldsymbol{\theta}^{(G)}$ and a local minimum $\delta^{(D)}$ with respect to $\boldsymbol{\theta}^{(D)}$. The solid lines represent information flow, whereas dotted lines show the flow of gradients. With a pretrained discriminator, the generator can generate $\boldsymbol{Y}$ in an unsupervised way.

Transformers are a type of neural network architecture that uses an attention mechanism to understand the relationships between the sequence of inputs ($\boldsymbol{X}$) and the sequence of outputs ($\boldsymbol{Y}$) by building up two neural network components – encoder and decoder[13] (Fig. 6). They can be pretrained on a large corpus of unlabelled data in a self-supervised fashion to gain background knowledge that improves performance on various natural language processing (NLP) jobs.

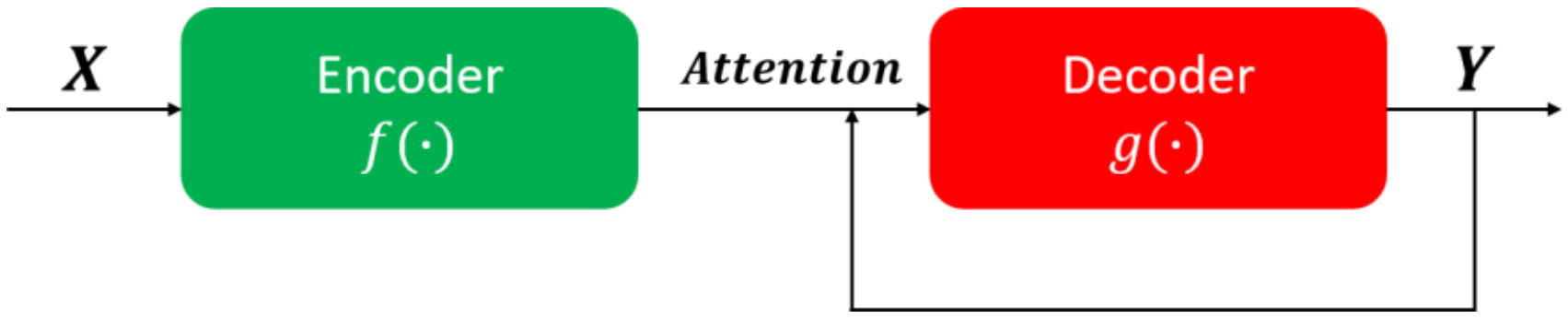

**Figure 6. Encoder-decoder architecture of transformers.** Transformers are a kind of seq2seq model with two components: an encoder, representing function $f(\cdot)$, which is a neural network with stacked multihead self-attention layers and other layers (feed-forward, normalization, etc.) transforming input sequences into context vectors, and a decoder, representing function $g(\cdot)$, which is a neural network with stacked multihead self-attention layers to receive preceding words and multihead attention layers to receive outputs from the encoder as well as from previous decoder blocks. Every output of the decoder depends on all words in the input sequence and all preceding outputs of the decoder, achieved by the attention mechanism. The gradients flow from decoder to encoder.

Regardless of which generative techniques are used, they either explicitly or implicitly model the distributions (likelihood functions) of datasets $\boldsymbol{X}$ and $\boldsymbol{Y}$, attempting to build a relationship between them via a set of functions $\boldsymbol{F}(\cdot)$, as shown in equation (4), in which the gradient of errors can be backpropagated through these functions in a chain of cascades to optimize the parameters of the functions towards minimizing errors of predictions.

$$\boldsymbol{Y} \leftarrow \boldsymbol{F}(\boldsymbol{X}; \boldsymbol{\theta}) \qquad (4)$$

Unlike supervised deep learning, in which $\boldsymbol{Y}$ is indicated by people (a form of intelligence), reinforcement learning and generative AI transform $\boldsymbol{X}$ to $\boldsymbol{Y}$ in an autonomous way by establishing relationships along time (Bellman backup in reinforcement learning and attention mechanism in transformers) and across space (sampling images from different scenes in GANs). Nevertheless, the chicken-or-egg problem reverberates around $\boldsymbol{X}$ and $\boldsymbol{Y}$, and what force is behind the transformations from $\boldsymbol{X}$ to $\boldsymbol{Y}$ and vice versa. This paper attempts to unveil this enigma in the following sections.

## Biological Neuronal Firing

The firing of bursts in neurons has been widely recognized to represent distinct functions or signal processing modes in biological brains, which relies on cellular mechanisms that work with feedback from higher centres to control the discharge properties of these cells[14]. The literature has also established computational models for high-level cognition based on biological mechanisms of the brain[15]. Although accurate models of the brain for distinctive aspects of human intelligence are still in quest, it is reasonable to abstract a biological brain as equation (1), while it is performing a specific intelligent task, and it is widely believed that the computational properties of biological neural networks are regulated by neural firing and neuromorphic at the molecular level in an efficient way[16,17,18], and the learning process of the brain is mediated by synaptic plasticity and neuromodulatory mechanisms[19]. While biological research in neuroscience has illuminated insights for the development of artificial intelligence[20], AI research has also shed light on understanding neural mechanisms[21].

We argue that a biological brain is a substrate to implement at the cortical or subcortical level a set of functions that map a series of inputs to outputs, as shown in equation (4), representing the intelligence of biological agents, whereas the molecular and cellular dynamics of neurons are to implement a set of modular functions ($\boldsymbol{\theta}$) compared with deep learning network architecture (Fig. 7). With this hypothesis, we further argue that a brain consists of many network components that are orchestrated together as the structures of deep neural networks, where gradient-descent and backpropagation akin mechanisms can be discovered for synaptic plasticity and neuromodulation.

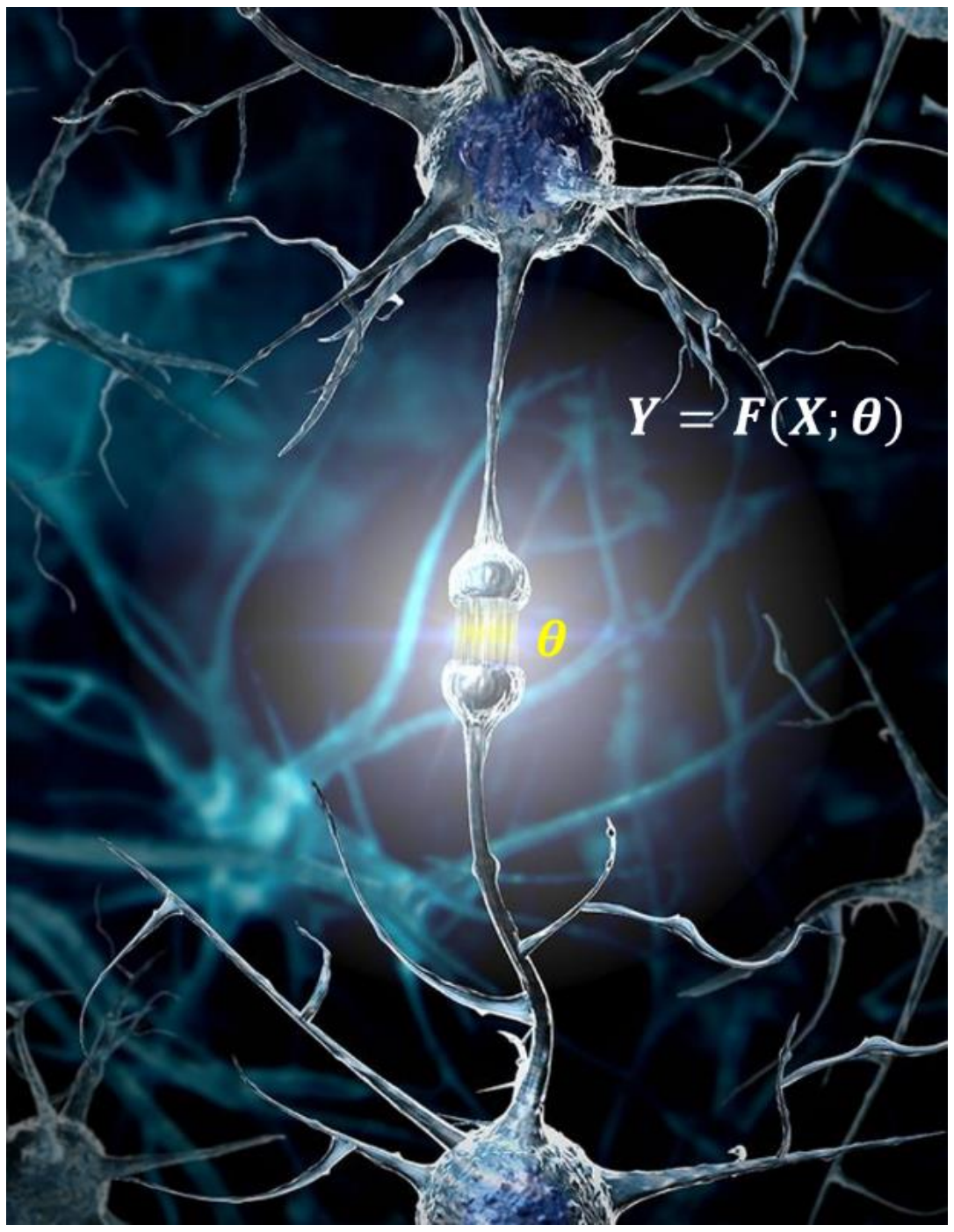

**Figure 7. Schematic illustration of biological neural networks representing a set of functions.** The cerebral cortex consists of large sets of neural networks that are primarily formed by neurons intricately connected with each other through dendrites and axons. Dendrites and axons form synapses where neurons activate other neurons via firing of bursts. The neuronal firings form the information flows from neurons to neurons conducted by the flow of inorganic ions and organic molecules, which is regulated by the cellular properties of synapses as well as the neuromodulators formulated in neural plasticity. All these biological properties discovered in neuroscience support that neural networks in the brain should represent or be represented by a set of functions ($\boldsymbol{F}(\cdot)$) mathematically, each of which is formulated by the composing synapses representing a series of modular functions ($\boldsymbol{\theta}$).

## Evolution of Life

As an efficient data storing and information processing mechanism, a gene and its expression system function with parameters optimized over evolution and rectified by natural selection. If representing genes as undependable variable $\boldsymbol{X}$ and proteins (in proper folding states for functioning) as dependable variable $\boldsymbol{Y}$, equation (2) is applicable for gene expression, and $\delta(\cdot)$ represents the natural selection process that favours the inherited mutations and extinguishes those that cannot survive on the Earth's environment. Thus, we hypothesize that the evolution of life on Earth follows the same path of machine learning, which maps available materials (nucleotides, amino acids, etc.) into complexes of functioning proteins (Fig. 8), and we argue that based on the progress of artificial intelligence, it can be verified by designing computational programmes that have recently shown initial successes in this area[22,23].

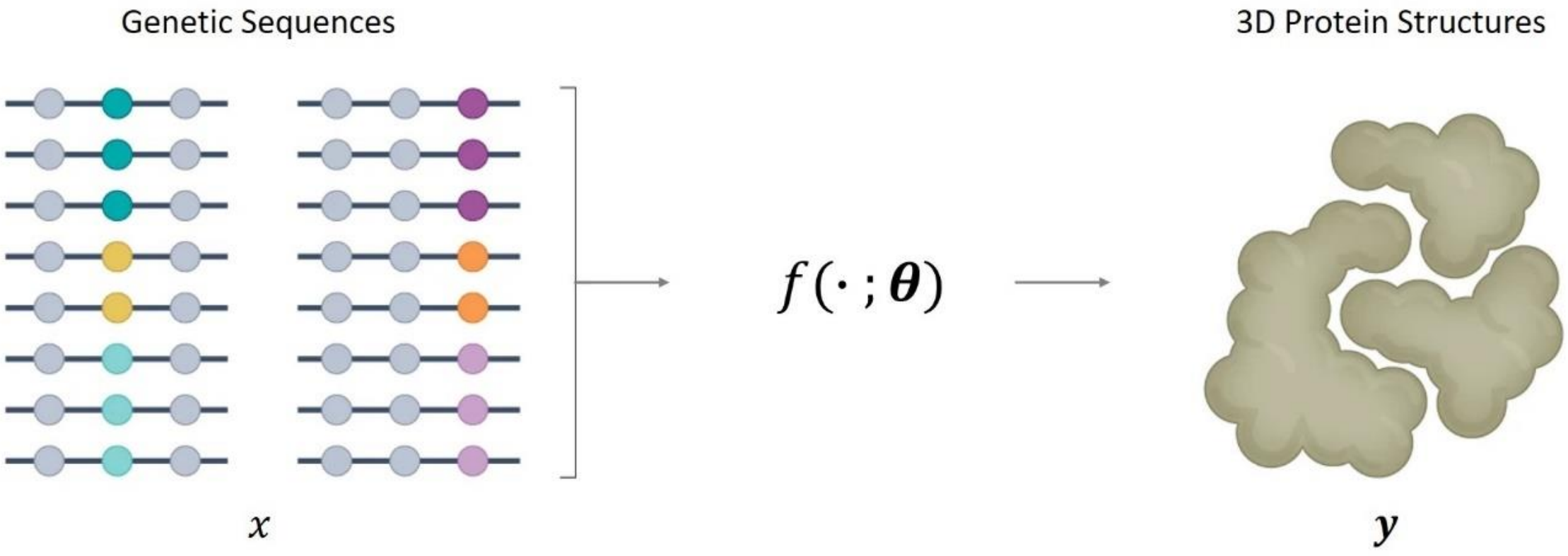

**Figure 8. Gene expression process in machine learning.** Representing genes as a dataset $\boldsymbol{X}$ and functioning proteins in proper 3D morphs as another dataset $\boldsymbol{Y}$, the gene expression process is an equation in which a function $f(\cdot)$ transforms $\boldsymbol{X}$ to $\boldsymbol{Y}$. With the development of AI, some models ($f(\cdot)$) with good results have been achieved.

## Human Intelligence and Unconsciousness

Based on the hypothesis of a brain representing a set of functions mapping inputs to outputs, we examine how humans respond to sensory stimuli before and after language appearance in history. In the early days of human history, people might countlessly meet wild animals and respond in different ways. For instance, when people encountered tigers, they might dodge or approach, and those who successfully avoided the danger survived, watching those who did not were killed by tigers, and experienced fearful feelings. These scenarios trained human brains by updating the modular parameters ($\boldsymbol{\theta}$) to minimize the errors. By minimizing the errors, people survived and kept the optimized parameters ($\boldsymbol{\theta}$), while those who did not were killed by ferocious animals. Gradually, humans build up a function $f(\cdot\,;\boldsymbol{\theta})$ in the brain to correctly respond to encountering different animals emotionally and behaviourally, with fear-and-dodge for fierce animals and calm-and-approach for amiable animals (Fig. 9).

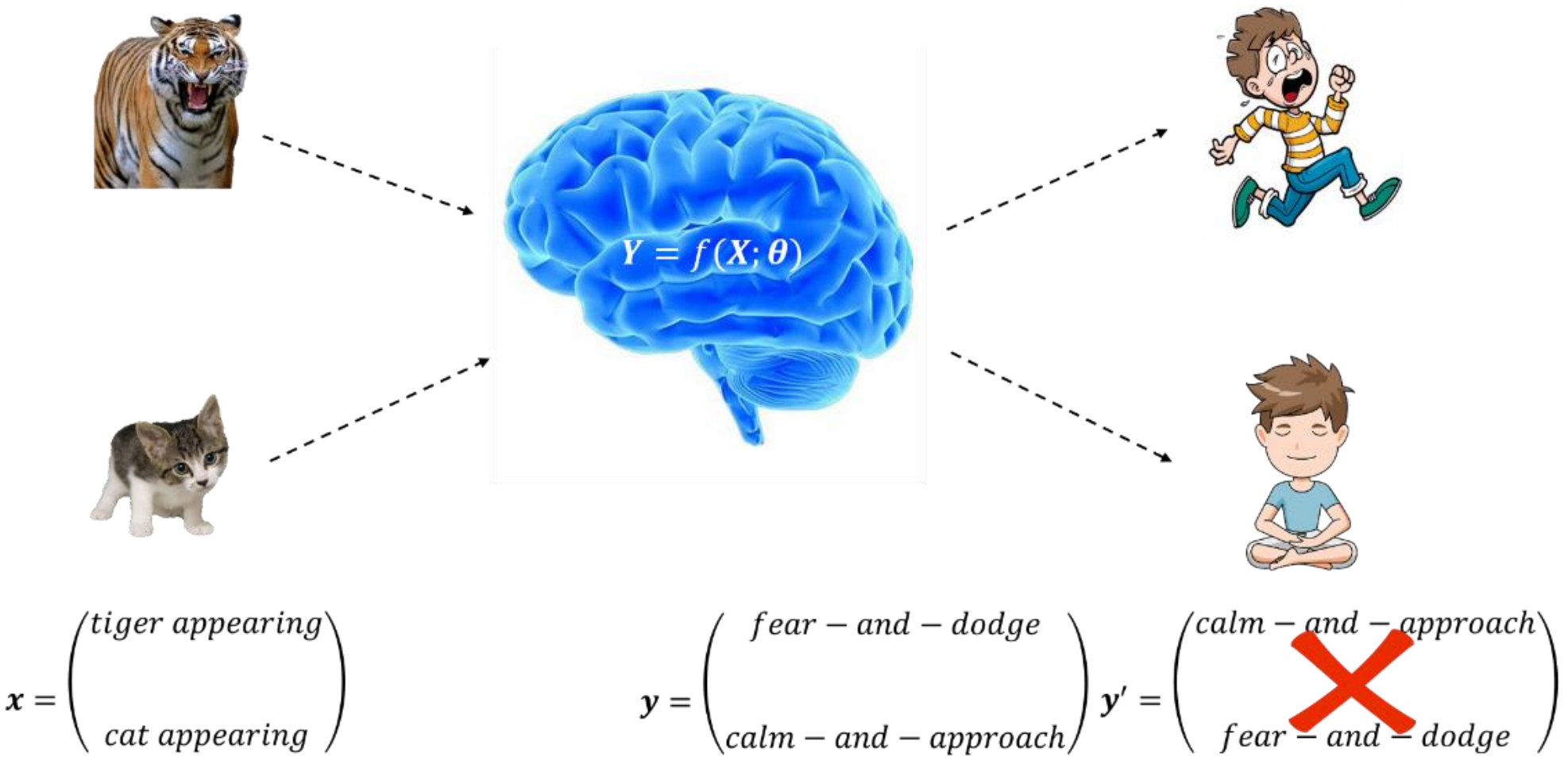

**Figure 9. Early humans learned to encounter animals.** Representing the decision mechanism responding to encountering animals in the human brain as a function $f(\cdot;\boldsymbol{\theta})$, early humans learn from countless encounters that form the samples ($\{\boldsymbol{x}\} \rightarrow \{\boldsymbol{y}\}$), and the parameters gradually converge to favour those mappings that survive humans and remove those that put humans in danger and extinguish them. For instance, when people meet tigers, they should be fearful and try to dodge, whereas they should keep calm and approach when meeting cats.

How the parameters ($\boldsymbol{\theta}$) are kept and passed down human evolution may rely on gene mutation, epigenetics[24,25,26,27,28,29] and cellular replication mechanisms. Before language appears, humans, similar to other animals, learn from trial-and-error methods by using sensory systems to perceive the environment ($\boldsymbol{X}$), determine behaviours and produce emotions ($\boldsymbol{Y}$), in which loss functions are defined by natural selection favouring behaviours to survive. It is reasonable to imagine that during the early ages of human history, a set of relatively stable mappings (equation 4) was formed, and people learned these mappings from generation to generation by observing and mimicking older ones. Older individuals use their own functions in the brain to generate samples of data, and younger generations are trained by these samples, in which the actual outputs $\boldsymbol{Y}$ are defined by older generations and a backpropagation akin mechanism is used to optimize the function parameters represented by cerebral structures of younger generations, minimizing the prediction errors. It is believed that emotion, as a special category of outputs, facilitates the calculation of prediction errors and updating the parameters of neuronal networks. Research in neuroscience has shed light on how emotion-relevant neuromodulators play the roles in this process[30,31].

The appearance of language is a large leap in human history that accelerates the learning process of humans about this world because it helps humans better describe the world as the special data ($\boldsymbol{X}$ and $\boldsymbol{Y}$). With language, samples are generated by tagging sensory inputs with words or lingual logics as the outputs, or even language represents the inputs themselves. Language plays an important role in training the brains of human individuals in a supervised way.

Summarizing the development of human intelligence, it occurs through three phases: 1) natural selection, in which the environment determines the cost function and minimizes errors by eliminating unfit behaviours while favouring fit behaviours; 2) mimicking, in which older generations define the actual outputs and younger generations learn from observing and imitating the elder's behaviours, in which the emotion facilitates signalling the actual outputs and prediction errors; and 3) language, in which the actual outputs are defined by language (words, sentences, stories) composed by other people and prediction errors are minimized by agreeing influential people (Fig. 10).

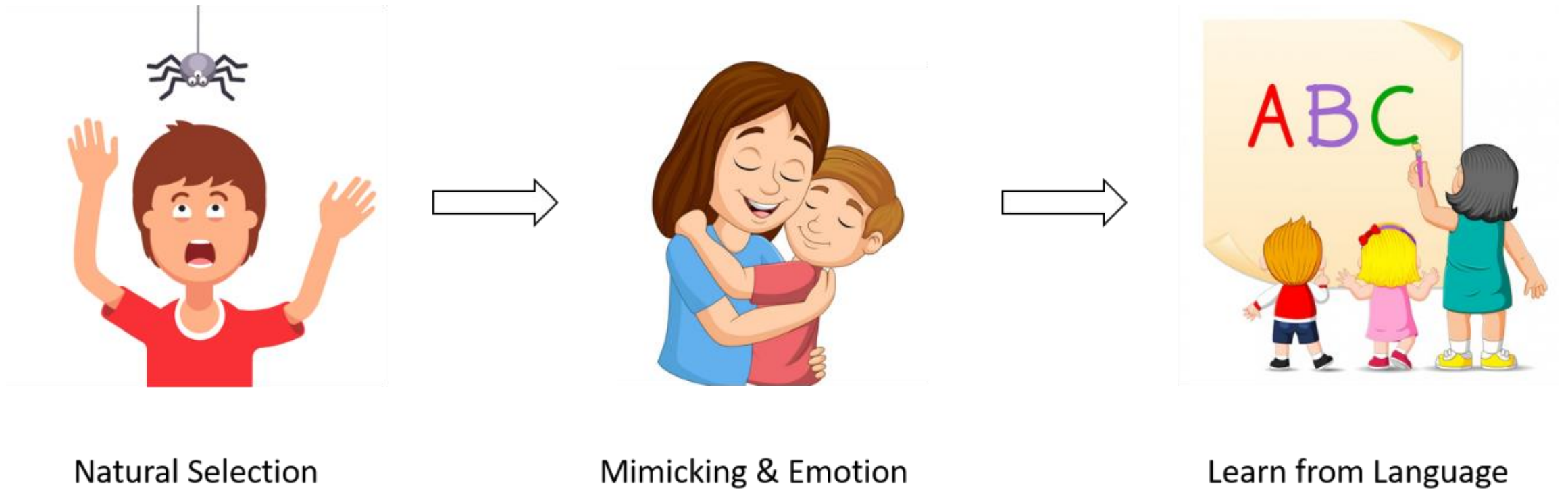

**Figure 10. Three phases of humans gaining intelligence along history.** In the early ages of human development, people primarily learn from interacting with the natural environment in a reinforcement-learning akin way, and natural selection plays an error-rectification role. Afterwards, humans living in families and communities learn from each other by mimicking others' behaviours, and in this process, emotional signals act as the main prediction error

representers. Last, after languages appear over human evolution, they are used as the most efficient tools for people to perceive this world and for annotating data sources into actual outputs of the intelligence-representing functions, and people's brains are trained by these annotations in a supervised way.

These phases can be defined as unsupervised, semi-supervised and supervised learning from the first to the third, respectively, according to modern AI nomenclature. These three learning processes can be formulated as equation (2). In the first phase, the human brain plays the function $f$, and the environment performs the error calculations as $\delta$, which rectifies the unfit behaviours to the environment by eliminating the hosts of outputs ($\boldsymbol{Y}$) that cannot survive. In this scenario, the destroyed hosts' parameters $\boldsymbol{\theta}$ will not be inherited along the evolution path, while those surviving keep and update their parameters via heredity, epigenetics and biological replication mechanisms. During the second phase, people learn from interacting with each other, mostly children from parents and other close ones. Stipulating $f$ represents a learner, and then the actual outputs ($\boldsymbol{Y}$) will be the imitated behaviours (including expression and emotion) that are acquired from order generations or from the first phase. In the last phase, the actual outputs ($\boldsymbol{Y}$) are indicated by other people via languages. All the learning processes of these three phases are ways in which biological brains copy their parameters. The first two phases determined common features of humans, endowing us with emotions, reflexes, basic desires and aversions, while languages divide humans into groups in which people have different values. However, intelligence obtained via equation (2) is essentially a series of modular functions and their parameters ($\boldsymbol{\theta}$), which automatically generates $\boldsymbol{Y}$ when encountering stimuli $\boldsymbol{X}$. This is called unconsciousness or subconsciousness.

## Language

It is reasonable to imagine that at the very beginning of language birth, symbols were generated randomly from a primitive neural network representing a function with arbitrary parameters ($\boldsymbol{\theta}$); therefore, one sensory input can be expressed as different symbols. Due to various reasons - for example, decay of recording material, fighting and killing, migration, and more to be verified by archaeology – the symbols converged to a stable set in an entropy-decreasing manner (equation 5). Whatever reason, a language system gradually formed, stored and evolved in one community, and language becomes one of the most important data formats for human intelligence.

$$\boldsymbol{Y}^*, \boldsymbol{F} = \arg\min_{\boldsymbol{Y},\boldsymbol{F}} H[\boldsymbol{Y} \leftarrow \boldsymbol{F}(\boldsymbol{X};\boldsymbol{\theta})] \qquad (5)$$

However, a key question arises when people learn from language – which annotation is the actual output $\boldsymbol{y}$ when people from different language communities face the same input $\boldsymbol{x}$? For instance, English-speaking people annotate a cat as 'Cat', whereas Chinese people map a cat to '猫' (Fig. 11). It is intelligent to map a real cat to 'Cat' by an English brain, similar to a Chinese brain, but when these two brains meet, there is no intelligence (they do not understand each other) until translational mapping is established between these two outputs, as shown in equation (6). Regardless of whether English is translated into Chinese or vice versa, we can measure whether there is an information entropy decline, and the decreased amount equals the mutual information of these two datasets (equation 7).

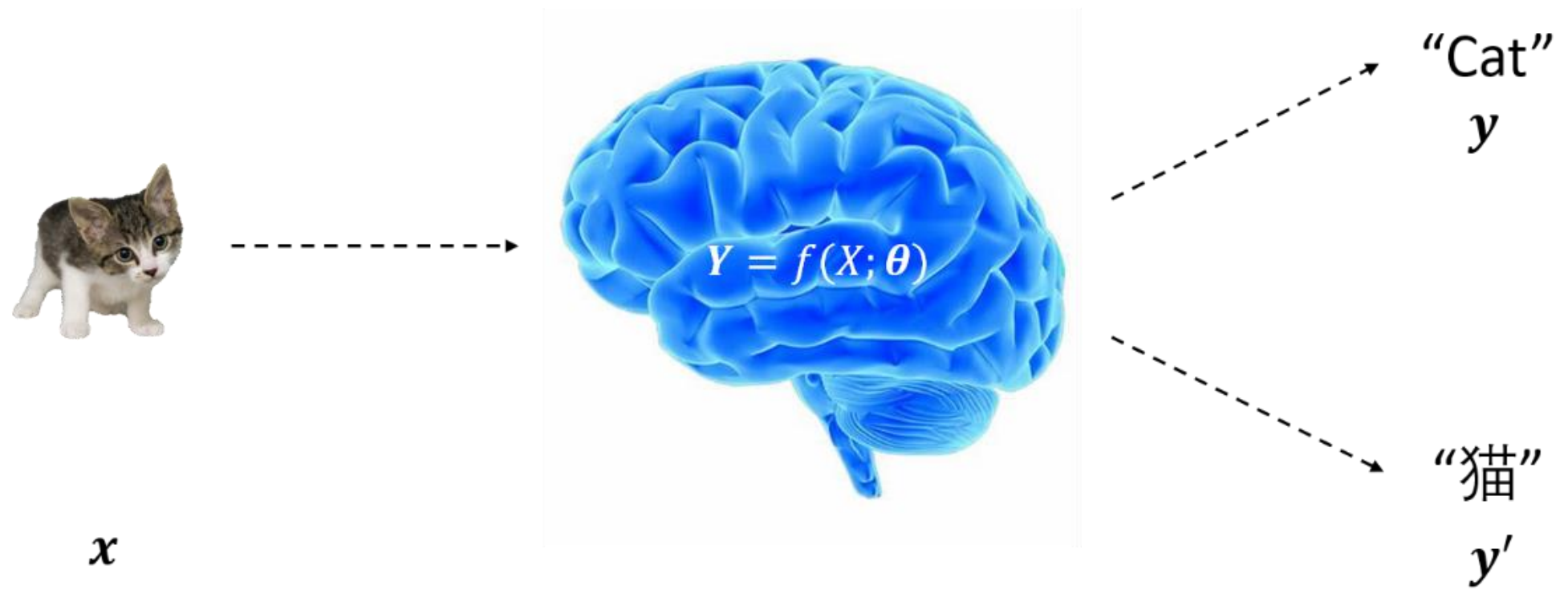

**Figure 11. The brain responds to two different language tags for the same input.** Representing a brain as a function $f(\cdot;\boldsymbol{\theta})$ when it is understanding a picture into a lingual descriptor, then the picture is a tensor value, and the outputs will be random variables representing words. This figure shows an example in which a cat can be transformed to an English word ‘Cat’ or a Chinese word ‘猫’, mediated by different function parameters $\boldsymbol{\theta}$.

$$\boldsymbol{Y}' \xleftrightarrow{f(\cdot\,;\,\boldsymbol{\theta})} \boldsymbol{Y} \qquad (6)$$

$$\begin{aligned} I(\boldsymbol{Y};\, \boldsymbol{Y}') &= H(\boldsymbol{Y}) - H(\boldsymbol{Y} \mid \boldsymbol{Y}') \\ &= H(\boldsymbol{Y}') - H(\boldsymbol{Y}' \mid \boldsymbol{Y}) \\ &= H(\boldsymbol{Y}) + H(\boldsymbol{Y}') - H(\boldsymbol{Y}, \boldsymbol{Y}') \end{aligned} \qquad (7)$$

By enclosing sensory information $\boldsymbol{X}$ and two different languages ($\boldsymbol{Y}$ and $\boldsymbol{Y'}$) into one system (Fig. 12), we analyse the process of the intelligence. Before language appearance, this world was there and perceived by humans as a dataset $\boldsymbol{X}$, with system entropy $H(\boldsymbol{X})$. After a language $\boldsymbol{Y}$ became stable within a community, a function was established in people’s brains to map sensory information into language. Then, the entropy of this system becomes $H(\boldsymbol{X},\boldsymbol{Y}) = H(\boldsymbol{X}) + H(\boldsymbol{Y}) - I(\boldsymbol{X};\boldsymbol{Y})$. If $\boldsymbol{X}$ and $\boldsymbol{Y}$ are independent of each other, then mutual information between them equals 0, and the entropy of this system never changes, in which case, there is no intelligence within this system until a dependent relationship is established by function $f_1$. This is the same for another community developing language $\boldsymbol{Y'}$ with function $f_2$. Now, these two communities established their own intelligence by using languages to understand this world, but when people from these two communities met, they cannot understand each other by speaking their own languages until a translational function $f_3$ was established in their brains to reduce the total entropy from $(H(\boldsymbol{Y}) + H(\boldsymbol{Y}'))$ to $H(\boldsymbol{Y},\boldsymbol{Y}')$ by mutual information $I(\boldsymbol{Y};\, \boldsymbol{Y}') > \boldsymbol{0}$. According to the property of joint entropy (equation 8), when the sensory information and two languages are independent of each other, the system has the largest entropy, while if the sensory information has deterministic relationships with both languages and two languages deterministically generate each other, then the system has the least entropy, equalling to the maximum marginal entropies of $\boldsymbol{X}$, $\boldsymbol{Y}$ and $\boldsymbol{Y'}$. Therefore, within this system, intelligence is a process that minimizes the information entropy of the system by establishing a series of functions mapping datasets in this system (Fig. 12).

$$\max[H(\boldsymbol{X}), H(\boldsymbol{Y}), H(\boldsymbol{Y}')] \le H(\boldsymbol{X},\boldsymbol{Y},\boldsymbol{Y}') \le H(\boldsymbol{X}) + H(\boldsymbol{Y}) + H(\boldsymbol{Y}') \qquad (8)$$

We argue that to implement more cognitive abilities, the human brain will develop more neural networks to represent more intelligent functions, and evidence favouring this hypothesis has been obtained by research on the structure of bilingual brains[32].

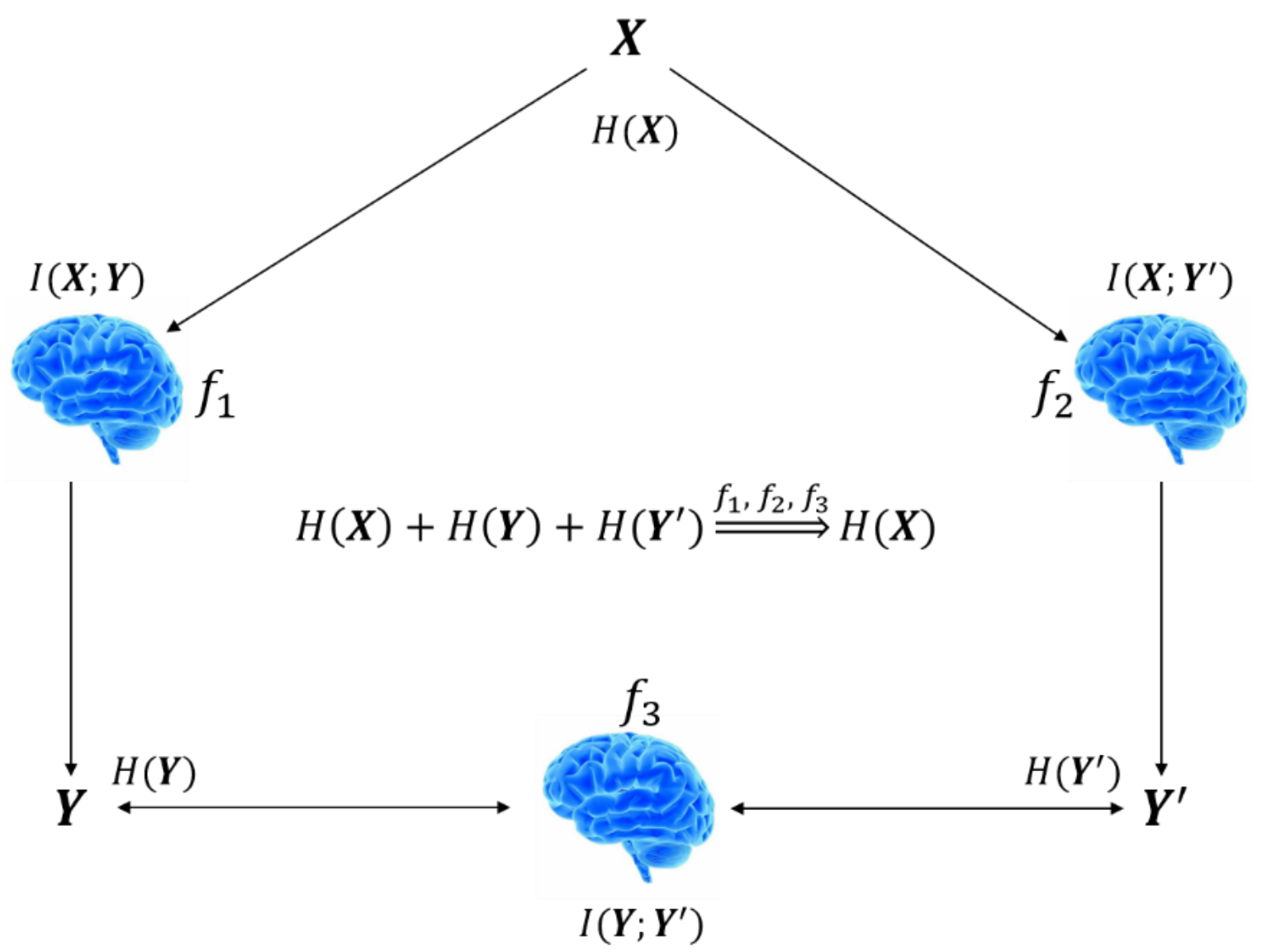

**Figure 12. Information analysis model for a bilingual intelligence system.** A bilingual brain maps the same data source $\boldsymbol{X}$ to two different outputs ($\boldsymbol{Y}$ and $\boldsymbol{Y}'$). If these two mappings are spatially separated, e.g., in two different brains, these two outputs have nothing about each other, indicating that there is no understanding between these two brains, and the total entropy of the system is $H(\boldsymbol{X},\boldsymbol{Y})+H(\boldsymbol{X},\boldsymbol{Y}')$. After a mapping connection is established between $\boldsymbol{Y}$ and $\boldsymbol{Y}'$ by a translator or growing their own translational neural network, they obtain full information about each other because they are both dependent on $\boldsymbol{X}$, and the system entropy is reduced towards $H(\boldsymbol{X})$ if $H(\boldsymbol{Y}|\boldsymbol{X})\to 0$ and $H(\boldsymbol{Y}|\boldsymbol{X})\to 0$, compared with the largest entropy value: $H(\boldsymbol{X})+H(\boldsymbol{Y})+H(\boldsymbol{Y}')$ in the case that three datasets are ideally independent of each other when there are no mapping functions between them.

However, language is a peculiar dataset that originates from symbolizing sensory stimuli and develops into an important stimulus sensed through visual and auditory systems (and the tactile system for braille) to generate more outputs. In the early development of language, sensory stimuli ($\boldsymbol{X}$) are mapped into a set of discrete symbols ($\boldsymbol{Y}_1=\{\boldsymbol{s}_i\}$) by the function $f_1$, which is an entropy-decreasing procedure that combines things into one space and time (spacetime), while the brain is a thermodynamical system that spontaneously scatters symbols over time to form a larger dataset ($\boldsymbol{Y}_2=\{\{\boldsymbol{s}_i\}_t\}$) with larger entropy (Fig. 13). However, brains tend to reduce system entropy by converting $\boldsymbol{Y}_2$ to another dataset $\boldsymbol{Y}_3$ with a similar structure to $\boldsymbol{Y}_2$. Depending on various compositions of symbols across space and time, the divergence between $\boldsymbol{Y}_2$ and $\boldsymbol{Y}_3$ oscillates above and under 0, and functions $f_2$ and $f_3$ will work when $(\boldsymbol{Y}_2-\boldsymbol{Y}_3)$ is positive and negative respectively. We define the divergence between the entropy of two datasets $\boldsymbol{X}$ and $\boldsymbol{Y}$ as the intelligence potential ($IP$) in equation (9). Functions $f_2$ and $f_3$ can be the same neural network, different networks, or even different layers of one neural network, which determines language context-based logics and syntax. Composed lingual expressions may be mapped between sensory stimuli by one or more functions with directions depending on the oscillation of the intelligence potential between these two datasets (Fig. 13).

$$IP_{XY}=H(\boldsymbol{X})-H(\boldsymbol{Y}), if\ 0<H(\boldsymbol{X}|\boldsymbol{Y})<H(\boldsymbol{X})\ and\ 0<H(\boldsymbol{Y}|\boldsymbol{X})<H(\boldsymbol{Y}) \quad (9)$$

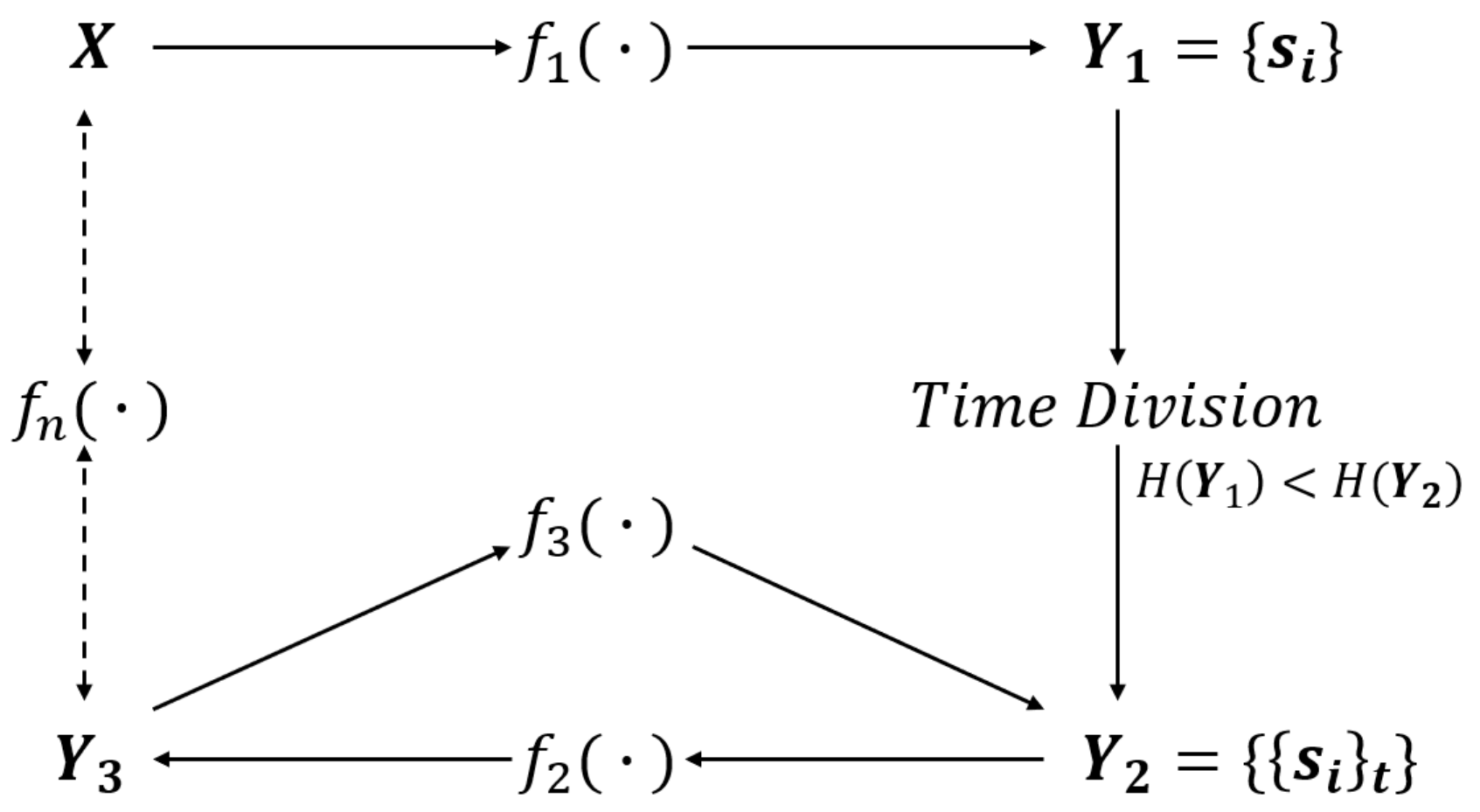

**Figure 13. Chained and looped information processing architecture of language neural networks.** In the early ages of human language development, the world ($\boldsymbol{X}$) is represented as scattered symbols ($\boldsymbol{Y}_1 = \{\boldsymbol{s}_i\}$) at one time horizon in the brain with entropy $H(\boldsymbol{Y}_1)$, but due to the spontaneous processes of entropy increasing, these symbols are divided over time into an expansive dataset $\boldsymbol{Y}_2 = \{\{\boldsymbol{s}_i\}_t\}$ with larger entropy ($H(\boldsymbol{Y}_2) > H(\boldsymbol{Y}_1)$). Brains tend to establish another function $f_2$ to transform $\boldsymbol{Y}_2$ into another time sequence $\boldsymbol{Y}_3$, but due to variations in time division, the relative divergence between $H(\boldsymbol{Y}_2)$ and $H(\boldsymbol{Y}_3)$ (we name it intelligence potential) will vibrate above and under zero. With the change in sign of the IP between positive and negative, brains tend to build functions to transform datasets from one and another in the direction of reducing entropy, and for the same reason, language sequences will be mapped back and forth to sensory stimuli about the world ($\boldsymbol{X}$). The vibration of IP should be a force behind the intelligent processes, and due to it, the language processing models have naturally chained and looped architecture, which determines language's context-based logics and syntax.

## Consciousness

Based on the above analysis, we argue that intelligence is a process of minimizing entropy by establishing a mapping function between datasets within a system, as shown in equation (10), where $\boldsymbol{\theta}^*$ is the optimal parameter of function $f(\cdot)$ approached through learning. To unify equation (2) and equation (10), we obtain that if there is a subset of outputs $\widetilde{\boldsymbol{Y}} \subseteq \boldsymbol{Y}$ to tag data for training, then the joint entropy minimizing process will be reduced to a process of minimizing errors $\delta(f(\boldsymbol{X}), \widetilde{\boldsymbol{Y}})$, and this is a process of gaining intelligence for a supervised-learning agent by mapping inputs onto a subset of real outputs (Fig. 14). This is the simplest and most efficient way of reducing the entropy of a system by mapping a dataset to a smaller subset, and it is similar to an information compression process by quantization. The three phases of human intelligence development all follow this process, as well as biological evolution by natural selection.

$$\boldsymbol{Y} \leftarrow f(\boldsymbol{X};\ \boldsymbol{\theta}): \boldsymbol{\theta}^* = \arg\min_{\boldsymbol{\theta}} H(\boldsymbol{X}, \boldsymbol{Y}) \qquad (10)$$

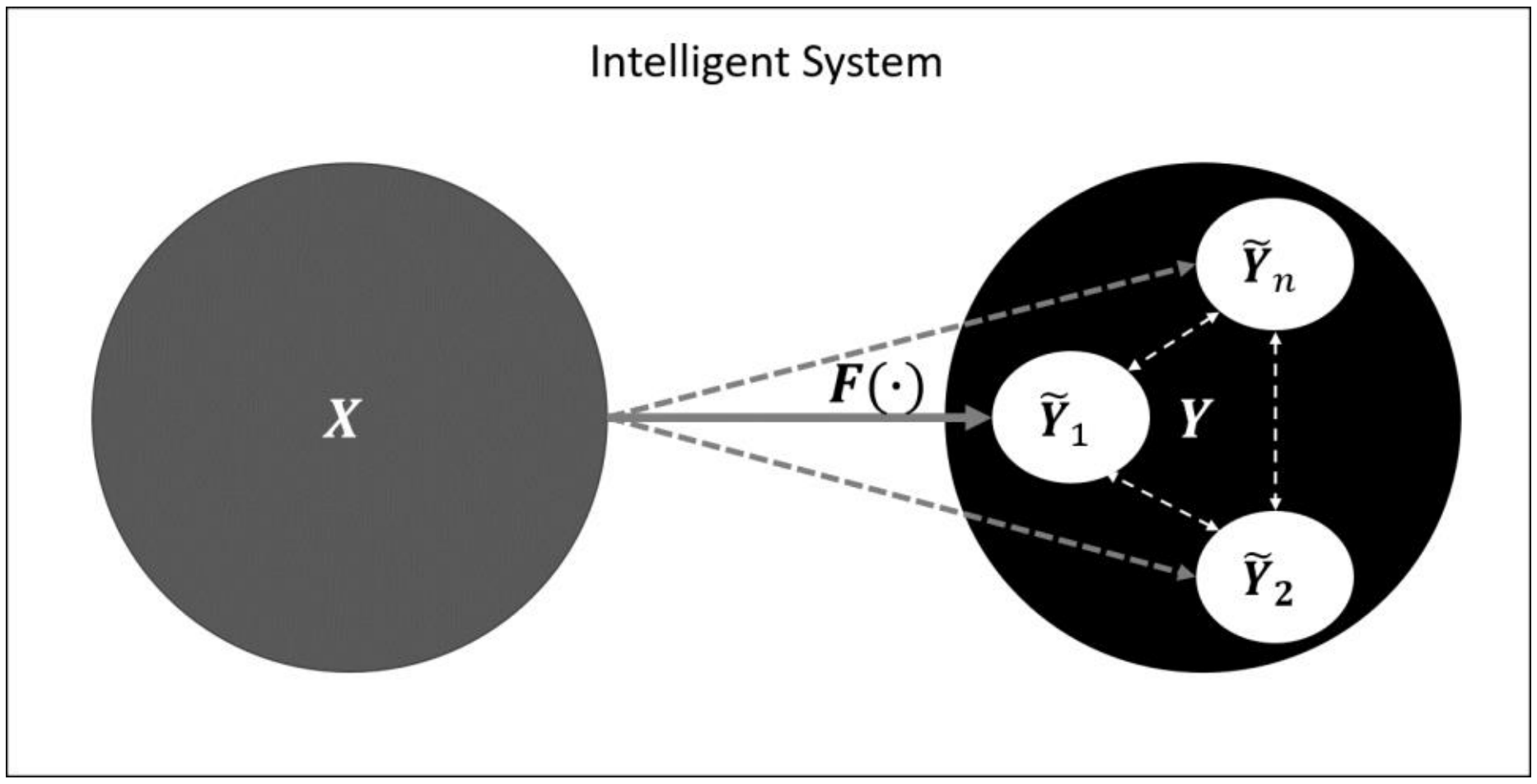

**Figure 14. An intelligent system to build unconsciousness and consciousness.** An intelligent system consists of two datasets $\boldsymbol{X}$ and $\boldsymbol{Y}$ and a set of functions between them. By mapping $\boldsymbol{X}$ to subsets of $\boldsymbol{Y}$, intelligent functions achieve the simplest and most efficient way of reducing system entropy on their own perspectives, which is the mechanism of supervised learning with annotated samples. However, the intelligent brains achieved in this way neither understand each other (not obtain any meaningful discerning from one another about the world $\boldsymbol{X}$) nor have any introspective comprehension of the world ($\boldsymbol{X}$). If the subsets ($\widetilde{\boldsymbol{Y}}_i$) are mapped from the same $\boldsymbol{X}$ by different brains living in the same society, they become conscious when they attempt to understand each other by building functional mappings between one another's outputs ($\widetilde{\boldsymbol{Y}}_i$), finally reaching a social consensus $\boldsymbol{Y}^*$ that mostly reflects the real ($\boldsymbol{X}$). This process gradually reduces the system entropy from $H(\boldsymbol{X}, \{\widetilde{\boldsymbol{Y}}_i\})$ to $H(\boldsymbol{X}, \boldsymbol{Y}^*)$, asymptotically approaching $H(\boldsymbol{X})$. The same process can happen in one brain scenario, in which perceptions from different sensory systems reach a consensus about the same stimuli. For various reasons, the set of different perceptions and the consensus in a society or a sensing brain keep changing as a dynamic process, so consciousness is also a dynamic process with entropy vibrations across neural networks. To track this dynamic process, a quantity called Consciousness Potential is defined in equation (16).

However, the subset mapping method for building intelligence is unconscious. How does human consciousness emerge? We hypothesize that consciousness emerges when the brain obtains consensus from different subsets of output by setting up connections between them (white arrows in Fig. 14). Imagining that there are $n$ brains that map $\boldsymbol{X}$ to $\widetilde{\boldsymbol{Y}}_1 \sim \widetilde{\boldsymbol{Y}}_n$ ($\widetilde{\boldsymbol{Y}}_i \subseteq \boldsymbol{Y}$), respectively, to every single brain, the system entropy is:

$$H(\boldsymbol{X}, \widetilde{\boldsymbol{Y}}_i) = H(\boldsymbol{X}|\widetilde{\boldsymbol{Y}}_i) + H(\widetilde{\boldsymbol{Y}}_i) \qquad (11)$$

When a brain tries to set up a connection from $\widetilde{\boldsymbol{Y}}_i$ to $\widetilde{\boldsymbol{Y}}_j$, it obtains information about $\widetilde{\boldsymbol{Y}}_i$ via $\widetilde{\boldsymbol{Y}}_j$:

$$I(\widetilde{\boldsymbol{Y}}_i; \widetilde{\boldsymbol{Y}}_j) = H(\widetilde{\boldsymbol{Y}}_i) - H(\widetilde{\boldsymbol{Y}}_i|\widetilde{\boldsymbol{Y}}_j) \qquad (12)$$

Plugging (11) into (12), we obtain:

$$I(\widetilde{\boldsymbol{Y}}_i; \widetilde{\boldsymbol{Y}}_j) = H(\boldsymbol{X}, \widetilde{\boldsymbol{Y}}_i) - H(\boldsymbol{X}|\widetilde{\boldsymbol{Y}}_i) - H(\widetilde{\boldsymbol{Y}}_i|\widetilde{\boldsymbol{Y}}_j) \qquad (13)$$

To maximize the information that a brain obtains from $\widetilde{\boldsymbol{Y}}_i$ about $\boldsymbol{X}$, according to equation (13), we obtain:

$$\begin{cases} H(\widetilde{\boldsymbol{Y}}_i|\widetilde{\boldsymbol{Y}}_j) \to 0 \\ H(\boldsymbol{X}|\widetilde{\boldsymbol{Y}}_i) \to 0 \end{cases} \quad (14)$$

Stipulating $\boldsymbol{Y}^*$ is the subset that minimizes $H(\boldsymbol{X}|\widetilde{\boldsymbol{Y}}_i)$ among $\widetilde{\boldsymbol{Y}}_i$ (get the most information from $\boldsymbol{X}$), brains in the same community tend to create a mapping function from $\boldsymbol{Y}^*$ to its original output $\widetilde{\boldsymbol{Y}}_i$ (original understanding about $\boldsymbol{X}$), as follows:

$$\widetilde{\boldsymbol{Y}}_i \leftarrow g(\boldsymbol{Y}^*, \boldsymbol{\mu}):\ \boldsymbol{\mu}^* = \underset{\boldsymbol{\mu}}{\operatorname{argmin}}\, H(\boldsymbol{Y}^*|\widetilde{\boldsymbol{Y}}_i) = \underset{\boldsymbol{\mu}}{\operatorname{argmin}}\, H(\boldsymbol{X}, \widetilde{\boldsymbol{Y}}_i) \quad (15)$$

In equation (15), $\boldsymbol{\mu}$ is the parameters of function $g(\cdot)$, and $\boldsymbol{\mu}^*$ is the optimal parameters obtained over learning. The dynamics of equation (15) represent the emergence of consciousness.

There are two kinds of consciousness: social consciousness and intrinsic consciousness. Individuals (not definitely human) acquire social consciousness by living in a society (community). Because every individual has different outputs (perception, understanding, emotion, behaviour, etc.) for the same world, they tend to obtain as much information as possible about this world by observing each other's behaviour and building interconnections, as shown in equation (13). In this process, those who know the most about the world ($\boldsymbol{X}$) win and define the social norm ($\boldsymbol{Y}^*$), and individuals develop neural networks in their brains to map $\boldsymbol{Y}^*$ to its original output $\widetilde{\boldsymbol{Y}}_{\boldsymbol{i}}$, which is mapped from $\boldsymbol{X}$ (Fig. 15). Because $\boldsymbol{Y}^*$ is changing in a society due to various reasons (such as changing demography), the social consciousness of individuals is a dynamic process.

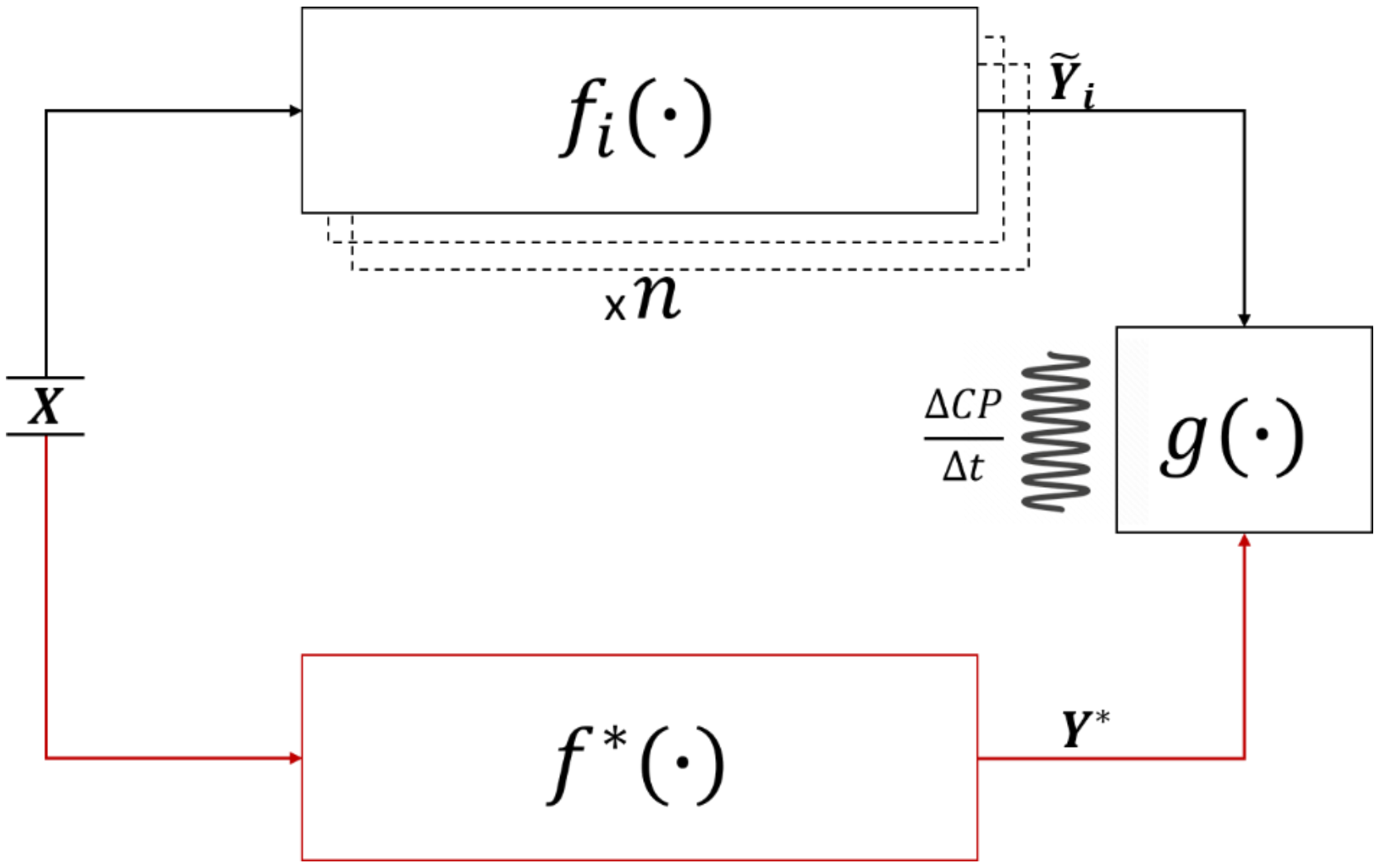

**Figure 15. Expansion of neural functions in parallel and semi-looped structures to generate consciousness.** The brain develops many neural networks $\{f_i(\cdot)\}$ in parallel for sensing stimuli as various outputs $\widetilde{\boldsymbol{Y}}_{\boldsymbol{i}}$. Due to relatively dependent entropies between the different outputs, brains tend to build neural networks $g(\cdot)$ mapping from one output to another in the direction of reducing system entropy. Because of natural changes in society and an individual's sensory systems, $g(\cdot)$ will keep changing the transformational directions directly or by using a looped structure, forming vibrations of consciousness potential (defined in equation 16) until CPs reach an equilibrium and a consensus among outputs $\widetilde{\boldsymbol{Y}}_{\boldsymbol{i}}$ is acquired as $\boldsymbol{Y}^*$. In this process, the pathway (in red) over neural networks $f^*(\cdot)$ is strengthened, and stimuli $\boldsymbol{X}$ tend to be comprehended by brains as $\boldsymbol{Y}^*$ (for human beings, this pathway is usually for language). This explains why people become unconscious about some stimuli after gaining social consciousness or intrinsic consciousness for a long time. However, with the changes in $\widetilde{\boldsymbol{Y}}_{\boldsymbol{i}}$ and $\boldsymbol{Y}^*$, the CPs over the neural networks $g(\cdot)$ keep vibrating, and individuals gain consciousness time over time.

It is imaginable that a person who lives in solitude from birth will not develop hope, intention and ideal-like social consciousness, which could be proven by examining the structural plasticity in its brain. For some pets who behaviourally develop consciousness while interacting with owners should have some expansions of subcortical structures to represent functions for social consciousness, compared with the wild counterparts.

In human society, $\boldsymbol{Y}^*$ can be social consensus defined by using language, such as constitution, morality and rules, and people who are knowledgeable have more influence over other people depending on how near they are approaching the real ($\boldsymbol{X}$). This influence is measured as a quantity defined by equation (16), which is called the consciousness potential ($CP$).

$$CP_{ij} = H(\boldsymbol{Y}_i|\boldsymbol{Y}_j) \geq 0,\ if\ H(\boldsymbol{X}|\boldsymbol{Y}_i) < H(\boldsymbol{X}|\boldsymbol{Y}_j) \qquad (16)$$

Do people who live alone can have consciousness? Answers are positive, and they develop intrinsic consciousness by developing neural networks to unify outputs from different sensory systems about the same stimuli. For example, some fruits are perceived by the visual system as being fresh and colourful, translated by the gustatory system to sweetness, smelled by the olfactory system as fragrance, and touched as being soft. All these outputs are associated by a neural network as ‘edible and delicious’ ($\boldsymbol{Y}^*$), that is, consciousness. Similar to social consciousness, intrinsic consciousness also has dynamic properties with oscillating consciousness potential due to reasons such as the development or damage of sensory systems that enhance, diminish or totally alter $CP$s. In Fig. 15, because of the vibration of $CP$ over time, the intelligence processes over $g$ may frequently change directions until reaching an equilibrium with a stable $CP$, in which a pathway over $f^*$ to obtain the most information about $\boldsymbol{X}$ is strengthened. In human brains, this strengthened pathway is usually for language, and due to being the major input for humans to perceive the world and being inputted from multiple sensory systems, language becomes the major carrier of human consciousness.

In summary, individuals interacting in social networks and multimodal neural networks are the basis of consciousness, and the dynamic processes of minimizing system entropy with consciousness potential oscillation over these networks push consciousness to emerge. This conforms to the neuroscience evidence that reveals that the unconsciousness depends on widespread neural networks where salient information gains access to consciousness when unconscious processing continues in parallel[33] and that the unconsciousness handles vastly more information than could reach our conscious minds at once[34].

## Entropy Conservation

Why does $\boldsymbol{X}$ need to be mapped to $\boldsymbol{Y}$? Which is the egg and which is the chicken? Is $\boldsymbol{Y}$ always generated from $\boldsymbol{X}$ or both exist before the connections are created by intelligence? To answer these questions, let us investigate the origin of life on Earth. One of the most significant matters on Earth is water ($H_2O$), which leads to the birth of life. Although there are many hypotheses for the source of water on Earth and the formation of the present planetary atmospheres[35], it is certain that water can be generated from combining both the primitive and basic elements of the universe: hydrogen ($H_2$) and oxygen ($O_2$), governed by the chemical equation (17). These two primitive gases may come from solar nebula or volcanic outgassing and helped build Earth’s primitive atmosphere[36,37].

$$2H_2O \leftarrow\ 2H_2 +\ O_2 \qquad (17)$$

While constraints remain on precise timing when equation (17) began to be applied on Earth or elsewhere, and although it is possibly due to photolysis for equation (17) to transpire in the opposite direction, it is widely believed that the existence of water on Earth is the most direct factor inducing life[38,39]. Why did hydrogen and oxygen meet to produce water on Earth, starting the history of life? Water is more structured at a fixed temperature and pressure and has less entropy by combining the same numbers of hydrogen ($H$) and oxygen ($O$), which collectively have more possibilities for arrangement and thus larger entropy as free gases. Collectively representing hydrogen ($H_2$) and oxygen ($O_2$) as variable $\boldsymbol{X}$, and water ($H_2O$) as $\boldsymbol{Y}$, equation (17) can be transformed to equation (1), in which function $f(\cdot\,;\boldsymbol{\theta})$ represents the condition and dynamics that hydrogen and oxygen react to become water, including the stoichiometry, which indicates the mass changes of molecules of individual components at the molecular level and the change in the amounts of components during reaction at the macroscopic level, the kinetics that reveals the time course of the reaction, and the conditions under which the reaction takes place or whether the reaction is completely unidirectional or leads to an equilibrium. Parameters $\boldsymbol{\theta}$ of function $f$ represent the environmental configurations with which the function can be realized, and $\boldsymbol{X}$ is transformed into $\boldsymbol{Y}$, such as the chemical spaces (the properties of other components present in the reaction mixture) for equation (17) to happen.

After water, the formation of early life on Earth can be formulated as equation (18), where $\boldsymbol{G}_x$ represents a group of prebiotic molecules and primitive biochemicals, and the equation establishes the environmental conditions conducted by water that are necessary for life.

$$\boldsymbol{EarlyLife} \leftarrow H_2O + \boldsymbol{G}_x \qquad (18)$$

No matter how complex the early lives were, a more orderly structure is achieved accompanying entropy reduction from the right side to the left side in equation (18). Life did not stop to develop towards increasing in complexity after it first appeared on Earth through evolutionary transitions[40,41]. Why does life evolve to increase in complexity over transitions? According to the second law of thermodynamics, the entropy of an isolated system always spontaneously increases in association with free energy dissipation. Does the evolution of life contradict the second law of thermodynamics? Maxwell coined a thought experiment in which an intelligent demon pumps heat from an isothermal environment and transforms it into work, decreasing the total entropy of the system. Maxwell's demon controversially violates the second law of thermodynamics, but it can be experimentally implemented to achieve Szilárd-type information-to-energy conversion[42].

Therefore, we hypothesize that in the universe, there are processes that decrease entropy by consuming energy, countering the spontaneous processes of the second law of thermodynamics, and on the whole, the total free energy released by spontaneous processes equals the total energy consumed by entropy-decreasing processes, which are intelligence. Because of the conservation of energy, the entropies decrease and increase in these opposite processes should counter each other, and the total entropy should also be conservative. In equation (19), the total accumulative entropy changes and energy transformation are both equal to zero, and an intelligence constant $C_{intl}$ is defined as the minus ratio of infinitesimal altered entropy to infinitesimal energy alteration, which measures the energy utilization efficiency by an intelligent system to reduce entropy.

$$\begin{cases} \oint dH = C_{intl} \oint dE = 0 \\ C_{intl} = -\dfrac{\partial H}{\partial E} \end{cases} \qquad (19)$$

Therefore, we answer why $\boldsymbol{X}$ needs to be mapped to $\boldsymbol{Y}$. That is, the intelligent processes countering spontaneous processes to keep the energy and entropy of the universe conservative. Based on this hypothesis, the appearance of water and the evolution of life on Earth all follow this intelligent process until the human brain appears, which deepens this process to human intelligence. From primitive gases to water, from water to life, from life to human brain, from human brain to abstract equations (Newton's law, theory of relativity etc.), all these processes not only reduce the entropy but also develop in a reinforced manner. How could these reinforced chains of functions happen? The equations established by intelligence keep generating $\boldsymbol{Y}$s from $\boldsymbol{X}$s, or just connect the originally existing $\boldsymbol{X}$s and $\boldsymbol{Y}$s? Just like the question we ask often about mathematics – is math invented or discovered? We argue that all are discovered. The nature of intelligence is to discover the relationships between datasets of the universe in an entropy-reducing way. The universe spontaneously develops towards disorder and releases energy by physically or informationally separating datasets with the space and time, but there are processes that move in opposite directions towards order by lifting these barriers and establishing connections between them to consume energy (Fig. 16). These opposite processes counter each other to keep the universe in equilibrium, conserving energy and entropy of the universe. In the universe, the curvature of the spacetime by gravity should be one of the major forces to lift the barriers.

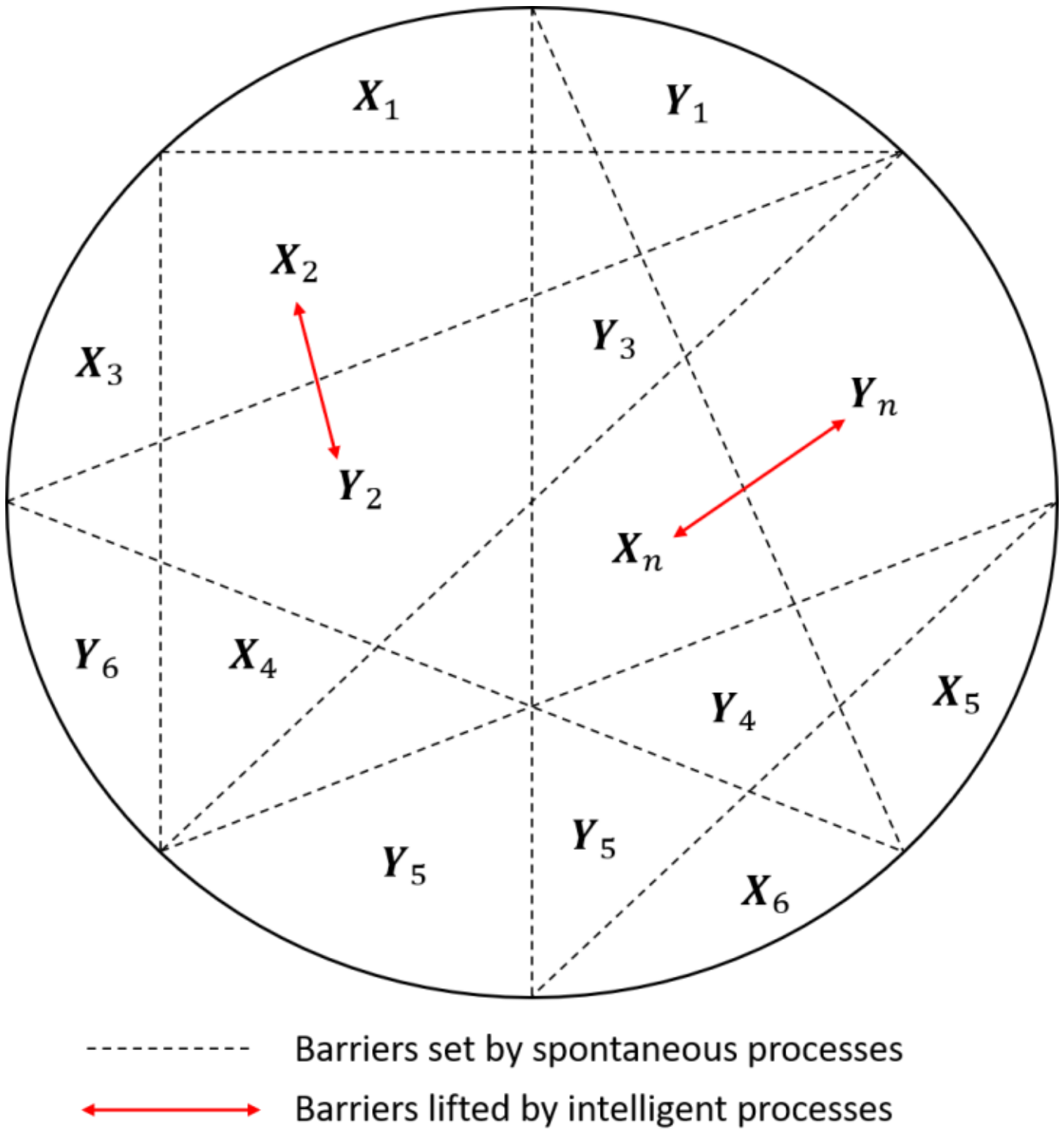

**Figure 16. Universe develops in two opposite directions to separate and merge datasets.** The universe is structured as space and time, which is called the spacetime by Einstein. Inside the universe, there are various datasets that are separated physically or informationally by spontaneous processes governed by the second law of thermodynamics to increase entropy, whereas an opposite force mediated by intelligent processes attempts to lift the barriers by establishing functional connections between the datasets. These two opposite processes, one of which is spontaneous, entropy-increasing and energy-releasing, and the other is entropy-reducing, energy-consuming and conducted by a kind of force (we call intelligence), counter each other to keep energy conservative and entropy conservative in the universe.

In Fig. 16, when an intelligent process establishes connections between two datasets, they are merged into one unified dataset, which increases the probability that it encounters other datasets in the space and time, favouring other intelligent processes to occur. That is why intelligence develops in a reinforced manner – from the appearance of water on Earth to life evolving in an accelerated way to today's human intelligence.

## Discussion

By reviewing artificial intelligence techniques, this research examines the evolution of life and human intelligence mechanisms, arguing that intelligence is a process in the universe that reduces entropy by setting up functional connections between datasets physically or informationally and concluding that the total entropy of the universe is conservative as well as energy. Furthermore, mathematical models for language, unconsciousness and consciousness are given in this paper before a qualitative relation between altered entropy and energy over the two counter processes of spontaneity and intelligence is suggested. The results of this paper predict that some sort of expansion of cortical plasticity could be discovered in animals that show consciousness over interacting with people or their peers in a kind of society, and AI agents can develop consciousness if they have a consensus mechanism established about outputs from different individuals' intelligent processes or from multimodal sensing of the environment. The hypothesis and conclusions of this paper can shed light on researches in neuroscience, psychology, physics and artificial intelligence. Predictions based on the hypotheses of this paper should be verified by experimental researches in relevant areas, and future research should be conducted to acquire quantitative formulas concerning the detailed development of intelligence in human and artificial agents, revealing the kinetics of the oscillation of intelligence potential ($IP$) and consciousness potential ($CP$) and their relations with gradient descent and back-propagation over neural networks.

With the conclusions of this paper, we further argue that human should not be the sole advanced intelligent agent in the universe. If AI or other forms of function representers can reduce entropy more efficiently in using energy, they can be more intelligent than us. However, no matter how accidentally human intelligence appears in this universe, we human beings are lucky to discover so many meaningful things, moving in the path of reducing the entropy of the universe. That is perhaps the meaning of life.

## Data availability

There are no data associated with this manuscript.

## Summary of notation

Capital letters are used for function sets and random variables, whereas lower case letters are used for functions and the values of random variables with bold font for tensors and normal font for scalars.

| | |
|---|---|
| $\leftarrow$ | assignment |
| $\rightarrow$ | asymptotic approach |
| $\boldsymbol{x} \in \boldsymbol{X}$ | $\boldsymbol{x}$ is one of the values of random variable $\boldsymbol{X}$ |
| $\widetilde{\boldsymbol{X}} \subseteq \boldsymbol{X}$ | $\widetilde{\boldsymbol{X}}$ is a subset of $\boldsymbol{X}$ or equals $\boldsymbol{X}$ |

$\underset{\boldsymbol{a}}{\operatorname{argmin}} f(\boldsymbol{a})$ a value of $\boldsymbol{a}$ at which $f(\boldsymbol{a})$ takes its minimum value

$\underset{\boldsymbol{a}}{\operatorname{argmax}} f(\boldsymbol{a})$ a value of $\boldsymbol{a}$ at which $f(\boldsymbol{a})$ takes its maximum value

$p(\boldsymbol{x})$ probability distribution of variable $\boldsymbol{X}$ at value $\boldsymbol{x}$

$\ln x$ natural logarithm of $x$

$H(\boldsymbol{X})$ entropy of random variable. Shannon's information entropy is used in this paper:

$$H(\boldsymbol{X}) = -\sum_i p(\boldsymbol{x}_i) \ln(p(\boldsymbol{x}_i))$$

$H(\boldsymbol{X}, \boldsymbol{Y})$ joint entropy of random variables $\boldsymbol{X}$ and $\boldsymbol{Y}$

$H(\boldsymbol{X}|\boldsymbol{Y})$ dependent entropy of $\boldsymbol{X}$ on $\boldsymbol{Y}$

$I(\boldsymbol{X}; \boldsymbol{Y})$ mutual information between random variables $\boldsymbol{X}$ and $\boldsymbol{Y}$

## Competing interests

The authors declare no competing interests.